\documentclass[aps,prd,floatfix,groupedaddress,nofootinbib,superscriptaddress,twocolumn]{revtex4-1}

\usepackage{amsmath,amssymb,bm,color,float,graphicx,mathrsfs}
\usepackage{hyperref}
\hypersetup{colorlinks=true,linkcolor=blue,citecolor=blue,urlcolor=blue}
\usepackage{Macro}

\def\T{\Theta}
\def\dtau{\delta\tau}
\def\taumean{\bar{\tau}}
\def\oT{\ol{\T}}
\def\src{\sigma}
\def\estt{\widehat{\dtau}}
\def\uestt{\ol{\dtau}}
\def\estp{\widehat{\phi}}
\def\ests{\widehat{\src}}
\newcommand{\NERSC}{{\tt NERSC}}
\newcommand{\NILC}{{\tt NILC}}
\newcommand{\SMICA}{{\tt SMICA}}
\newcommand{\MILCA}{{\tt MILCA}}

\def\Cty{C_L^{\tau y}}
\def\Ctt{C_L^{\tau \tau}}

\newcommand{\uvec}[1]{\hat{\boldsymbol{#1}}}

\usepackage{xcolor}

\begin{document}
\title{Constraining reionization with the first measurement of the cross-correlation between the CMB optical-depth fluctuations and the Compton y-map}

\author{Toshiya Namikawa}
\affiliation{Department of Applied Mathematics and Theoretical Physics, University of Cambridge, Wilberforce Road, Cambridge CB3 0WA, United Kingdom}

\author{Anirban Roy}
\affiliation{Department of Astronomy, Cornell University,\\ Ithaca, New York 14853, USA}

\author{Blake D. Sherwin}
\affiliation{Department of Applied Mathematics and Theoretical Physics, University of Cambridge, Wilberforce Road, Cambridge CB3 0WA, Unite Kingdom}
\affiliation{Kavli Institute for Cosmology, University of Cambridge, Madingley Road, Cambridge CB3 OHA, United Kingdom}

\author{Nicholas Battaglia}
\affiliation{Department of Astronomy, Cornell University,\\ Ithaca, New York 14853, USA}

\author{David N. Spergel}
\affiliation{Center for Computational Astrophysics, Flatiron Institute,\\ 162 Fifth Avenue, New York, New York 10010, USA}
\affiliation{Department of Astrophysical Sciences, Princeton University, Peyton Hall, Princeton, New Jersey 08544, USA}


\date{\today}

\begin{abstract}
We propose a new reionization probe that uses cosmic microwave background (CMB) observations; the cross-correlation between fluctuations in the CMB optical depth which probes the integrated electron density, $\delta\tau$, and the Compton $y$-map which probes the integrated electron pressure. This cross-correlation is much less contaminated than the $y$-map power spectrum by late-time cluster contributions. In addition, this cross-correlation can constrain the temperature of ionized bubbles while the optical-depth fluctuations and kinetic SZ effect can not. We measure this new observable using a Planck $y$-map as well as a map of optical-depth fluctuations that we reconstruct from Planck CMB temperature data. We use our measurements to derive a first CMB-only upper limit on the temperature inside ionized bubbles, $T_{\rm b}\lesssim 7.0\times10^5\,$K ($2\,\sigma$). We also present future forecasts, assuming a fiducial model with characteristic reionization bubble size $R_{\rm b}=5\,$Mpc and $T_{\rm b}=5\times10^4\,$K. The signal-to-noise ratio of the fiducial cross-correlation using a signal dominated $y$-map from the Probe of Inflation and Cosmic Origins (PICO) becomes $\simeq7$ with CMB-S4 $\delta\tau$ and $\simeq13$ with CMB-HD $\delta\tau$. For the fiducial model, we predict that the CMB-HD $-$ PICO cross-correlation should achieve an accurate measurement of the reionization parameters; $T_{\rm b}\simeq 49800^{+4500}_{-5100}\,$K and $R_{\rm b}\simeq 5.09^{+0.66}_{-0.79}\,$Mpc. Since the power spectrum of the electron density fluctuations is constrained by the $\delta\tau$ auto spectrum, the temperature constraints should be only weakly model-dependent on the details of the electron distributions and should be statistically representative of the temperature in ionized bubbles during reionization. This cross-correlation could, therefore, become an important observable for future CMB experiments.
\end{abstract} 



\maketitle


\section{Introduction}

Several hundred million years after the hot big bang the universe was reionized. This process is poorly understood and there are many open questions about the reionization epoch: When did reionization start? How long did it last? What objects reionized the universe? Was this a smooth or highly inhomogenuous process?

There are many different observational constraints on the timing and duration of the reionization epoch. 
Observations of the Ly$\alpha$ absorption in high-redshift quasar spectra indicate that reionization had completed between $z=5$ and $6$ \cite{McGreer:2014,GunnPeterson:1965,Kulkarni:2018}. 
The CMB optical depth, which is induced by the free electrons along the line of sight, has been constrained by the angular power spectrum of CMB anisotropies \cite{P16:reion, P18:main}. 
Recent measurements of the CMB temperature and polarization spectra \cite{P16:reion} suggest that the ionization fraction at $z\agt 10$ is less than $10\%$, so that an early onset of reionization above is highly disfavored. 

In contrast, constraints on the properties of ionized bubbles such as their temperature and size are still very poor. 
During the reionization epoch the ionized bubbles in the intergalactic medium were formed by radiation from the first star-forming galaxies and their sizes grew toward the end of reionization. The ionized bubbles were heated significantly with their temperature eventually rising to $\sim 10^4$\,K \cite{McQuinn:2016}. Constraints on the properties of the ionized bubbles can therefore provide insights into the mechanism of cosmic reionization.

An inhomogeneous reionization process produces significant spatial fluctuations of the electron density at the reionization epoch which lead to anisotropies in the optical depth of the CMB. The anisotropy depends on the reionization history, the characteristic radius of ionized bubbles, and their distribution \cite{Wang:2006, Dvorkin:2008:tau-est}. The anisotropies of the CMB optical depth have been constrained by several analyses \cite{Gluscevic:2012:tau,Namikawa:2017:plktau,Feng:2018:tau-phi}. 
However, the measurements so far have been limited by significant reconstruction noise. Assuming the analytic model of \cite{Wang:2006}, measurements of the angular power spectrum of the optical depth have put an upper bound on the characteristic bubble size as $R_{\rm b}\alt 10$~Mpc at $2\,\sigma$ \cite{Namikawa:2017:plktau}, while theoretical and simulation-based studies \cite{Furlanetto:2006,Zahn:2006b, Battaglia:2013} suggest $R_{\rm b}\sim\mC{O}(1-10)$\,Mpc. 
Future experiments such as CMB-S4 as well as future 21-cm observations such as the Square Kilometre Array (SKA) will be able to constrain the bubble size down 
to $\sim 1$\,Mpc in the redshift range $6-10$ \cite{Wyithe:2004, Mellema:2015}. \footnote{We note that the recent measurement of the global 21cm signal
from reionization by EDGES could, if confirmed, indicate that the true reionization process significantly deviates from the current standard scenario of reionization
\cite{Bowman:2018}.}
Progress in such observations of inhomogeneous reionization will allow us to test our current knowledge of reionization and gain new insights into the reionization epoch. 

The Sunyaev Zel'dvich (SZ) effect generated by hot electrons during the reionization epoch can be also used to explore the reionization epoch. In particular, the thermal SZ (tSZ) effect is sensitive to the temperature of the ionized bubbles, which makes the tSZ signal a unique probe of the gas temperature in the reionization era. However, a challenge is that the tSZ effect is dominated by the late-time contributions from clusters \cite{Hill:2015}. Ref.~\cite{Baxter:2020dvd} proposes to use the cross-correlation between tSZ and high-redshift galaxies for another reionization probe. Alternatively, the kinetic SZ (kSZ) effect generated by the radial motion of ionized bubbles during reionization could be detected by future CMB experiments, which will potentially constrain the redshift and duration of reionization \cite{Mesinger2012,Battaglia2013b,Smith:2016lnt, Alvarez:2020}. However, the kSZ effect does not directly constrain the temperature of the ionized bubbles. 

In this work, we propose a new approach to probe reionization via the cross-correlation of the anisotropies of the CMB optical depth and the tSZ effect. The analysis of this cross spectrum has the following advantages: the cross-correlation can constrain the temperature of the ionized bubbles, but the late-time cluster tSZ contribution to the cross-correlation is less significant than for a tSZ measurement alone. 

We apply this cross-correlation method to Planck data and derive constraints on reionization from our measurement; we also perform forecasts for future experiments with this method.Throughout this work we assume a flat $\Lambda$CDM universe with the cosmological parameters defined by \textit{Planck} TT, TE, EE+lowE+lensing \cite{P18:main}.

\section{Theoretical model} \label{sec:theory}

The change in temperature at each frequency $\nu$ due to the tSZ effect is expressed as:
\al{
    \frac{\Delta T_{\nu}}{T_{\rm CMB}} = g(\nu)y \,, 
}
where $g(\nu)=x\coth(x/2)-4$ with $x=h\nu/(k_{\rm B}T_{\rm CMB})$ ignoring the relativistic corrections, $T_{\rm CMB}$ is the CMB temperature, $k_{\rm B}$ is the Boltzmann constant, $h$ is the Planck constant, and $y$ is the Compton-$y$ parameter.

Temperature anisotropies due to the tSZ effect have contributions from both reionization and clusters. 
Hot electron gas inside the ionized bubbles is parametrized either by the electron gas temperature (temperature inside ionized bubbles), $T_{\rm b}$, and the electron number density, $n_{\rm e}$, separately, or by the pressure, $P_{\rm e}=k_{\rm B}n_{\rm e}T_{\rm b}$. The Compton-$y$ parameter is given as: 
\al{
    y(\uvec{n}) = \frac{\sigma_{\rm T}}{m_{\rm e}c^2}
    \Int{}{\chi}{}ak_{\rm B} T_{\rm b}(\uvec{n},\chi) n_{\rm e}(\uvec{n},\chi) 
    \,, \label{eq:y_reio}
}
where $\uvec{n}$ is the unit vector along the line of sight, $m_{\rm e}$ is the electron mass, $\sigma_{\rm T}$ is the Thomson scattering cross-section, 
$c$ is the speed of the light in free space, $a$ is the scale factor and $\chi$ is the comoving distance. The ionization fraction, $x_{\rm e}$, traces the phase transition from neutral to ionized states of the Universe, 
and is defined as the ratio of the free electron density to the neutral hydrogen density. In particular, if $n_{\rm p0}$ is the density of protons at present,
we have $n_{\rm e}=n_{\rm p0}x_{\rm e}/a^3$. 
Thus, the $y$ parameter depends on both the electron temperature, $T_{\rm b}$, and ionization fraction, $x_{\rm e}$.

The optical depth measures the integrated electron density along the line of sight: 
\al{
\tau(\uvec{n})=\sigma_{\rm T}\Int{}{\chi}{} an_{\rm e}(\uvec{n},z)
\,. 
}
If the temperature of ionized regions in any specific direction $\uvec{n}$ is constant during the epoch of reionization (EoR), $y$ could be determined through the optical depth as $y=(k_{\rm B} T_{\rm b}/m_{\rm e}c^2)\times \tau \simeq 9.1\times 10^{-8}(\tau/0.054)(T_{\rm b}/10^4 \rm K)$ \cite{DeZotti:2015}.

Denoting the fluctuation part of $\tau$ as $\dtau$, the angular cross-spectrum between $\dtau$ and $y$ due to inhomogeneous reionization, $\Cty$, is given under the Limber approximation as:
\al{
    \Cty = \frac{k_{\rm B}\sigma_{\rm T}^2n_{\rm p0}^2}{m_{\rm e}c^2}\Int{}{\chi}{a^4\chi^2}T_{\rm b}(\chi)P_{x_{\rm e}x_{\rm e}}\left(k=\frac{L+1/2}{\chi},\chi\right)
    \,, \label{eq:cltauy}
}
and the angular power spectra of $\delta\tau$ can be expressed as:
\al{
    \Ctt = \sigma_{\rm T}^2 {n}^2_{\rm p0}\Int{}{\chi}{a^4\chi^2} 
    P_{x_{\rm e}x_{\rm e}}\left(k=\frac{L+1/2}{\chi}, \chi\right) 
    \,. \label{eq:tautau}
}
In the above equations, $P_{x_{\rm e}x_{\rm e}}$ is the three-dimensional power spectrum of density-weighted ionization fraction fluctuations at comoving distance $\chi$. This quantity captures the physics and morphology of reionization, and in particular, the size and distribution of ionized bubbles. As reionization is a complex and poorly constrained process, it is a difficult task to model the distribution of free electrons in different redshift bins during EoR. Therefore, we use the halo model approach to model $P_{x_{\rm e}x_{\rm e}}$ given by \cite{Wang:2006} in which the ionized bubbles are associated with dark matter halos. We refer the reader to \cite{Wang:2006, Mortonson:2006:model, Dvorkin:2008:tau-est, Roy:2018} for further details on the modeling of $P_{x_{\rm e}x_{\rm e}}$ and \cite{Roy:2020} for comparison with simulations. 

In modeling $P_{x_{\rm e}x_{\rm e}}$, we need a size distribution of the ionized bubbles. To see this, we first note that the model describes the probability that a given point in space, $\Vec{r}$, is ionized as a Poisson process:
\al{
    \ave{x_{\rm e}(\Vec{r})}_{\rm P} = 1-\exp(-n_{\rm b}(\Vec{r}) V_{\rm b}) 
    \,,
}
where the brackets, $\ave{\cdots}_{\rm P}$, denote averaging over the Poisson process and $n_b$ is the number density of ionized bubbles. $V_{\rm b}$ is the volume of ionized bubbles which is given by \cite{Mortonson:2006:model}:
\al{
    V_{\rm b} \equiv \Int{}{R}{}\frac{4\pi R^3}{3}P(R) \,,
}
where $P(R)$ is a size distribution of the ionized bubbles. We assume the following log-normal distribution for the bubble size distribution \cite{Mortonson:2006:model,Dvorkin:2008:tau-est}:
\al{
    P(R) = \frac{1}{R}\frac{1}{\sqrt{2\pi\sigma_{\rm lnr}^2}}\exp{\left\{-\frac{[\ln\left(R/{R_{\rm b}}\right)]^2}{2\sigma_{\rm lnr}^2}\right\}} 
    \,. \label{Eq:pdf-R}
}
Here, $R_{\rm b}$ is the characteristic radius of ionized bubbles and $\sigma_{\rm lnr}$ is the width of the log-normal distribution. These parameters are in general a function of $z$, while radiative transfer simulations are required to evaluate such $z$-distribution. Following the previous studies \cite{Mortonson:2006:model,Dvorkin:2008:tau-est}, we assume that $R_{\rm b}$ and $\sigma_{\rm lnr}$ are independent of $z$. 

Note that we ignore the perturbations to the electron temperature throughout this paper. The fluctuations of $T_{\rm b}$ could be generated, for example, if the temperature fluctuations are induced by the gas density fluctuations which produce correlations between $\ave{T_{\rm b}x_{\rm e}x_{\rm e}}$ in $\Cty$. If the fluctuations of the ionization fraction also trace the gas density fluctuations, this correlation could enhance $\Cty$ and tighten the temperature constraint. We provide a quantitative analysis on the temperature fluctuations in Appendix \ref{app:Te} assuming a small gas density perturbations and find that the correction term only enhances $\Cty$ by several percent. An accurate modeling of $\Cty$ in the presence of the temperature fluctuations, however, requires a radiative transfer simulation because the reionization process would be non-Gaussian and gas density perturbations could be large. We defer a detailed study on the impact of the temperature fluctuations on $\Cty$ to future work. 

\section{Method} \label{sec:method}

The $y$-map can be extracted from multifrequency temperature maps via the frequency dependence of the tSZ effect. We use the $y$-maps provided by Planck, which are based on internal linear combination techniques such as the Needlet Internal Linear Combination ({\tt NILC}; \cite{Remazeilles:2011:NILC}) and the Modified Internal Linear Combination Algorithm ({\tt MILCA}; \cite{Hurier:2013:MILCA}). 

We reconstruct the optical depth anisotropies as follows. The non-zero optical depth leads to a suppression of small-scale temperature fluctuations by $\E^{-\tau}$. Denoting the temperature fluctuations in the absence of $\delta\tau$ as $\T$, the observed temperature anisotropies become:
\al{
	\hT(\hatn) &= \E^{-\dtau(\hatn)}\T(\hatn) 
		= \T(\hatn) - \dtau(\hatn)\T(\hatn) + \mC{O}(\dtau^2) \,.
}
The second term leads to mode coupling between the temperature anisotropies ($\l\not=\l',m\not=m'$) \cite{Dvorkin:2008:tau-est}: 
\al{
    \ave{\hT_{\l m}\hT_{\l'm'}}_{\rm CMB} 
    = \sum_{LM}\Wjm{\l}{\l'}{L}{m}{m'}{M} f^\tau_{\l L\l'}\dtau^*_{LM}
    \,, \label{Eq:tau-mixing}
}
where $\ave{\cdots}_{\rm CMB}$ denotes an ensemble average where the realization of the optical depth anisotropies is held fixed and $f^{\dtau}_{\l L\l'}$ is defined as:
\al{
	f^{\dtau}_{\l L\l'} &= - (\CTT_\l + \CTT_{\l'})
	\gamma_{\l L\l'}\Wjm{\l}{L}{\l'}{0}{0}{0}
	\,, \label{Eq:weight:tau}
}
with $\CTT_\l$ being the temperature power spectrum and 
$\gamma_{\l L\l'}=[(2\l+1)(2L+1)(2\l'+1)/4\pi]^{1/2}$. 
From \eq{Eq:tau-mixing}, we can derive a quadratic estimator to reconstruct the optical-depth anisotropies expressed as a convolution of two temperature fluctuations with appropriate weights; this process is very similar to CMB lensing reconstruction \cite{Dvorkin:2008:tau-est}. 

We first compute the following unnormalized estimator: 
\al{
	\uestt^*_{LM} = \frac{1}{2}\sum_{\l m\l'm'} \Wjm{\l}{\l'}{L}{m}{m'}{M} 
	f^{\dtau}_{\l L\l'} \oT_{\l m}\oT_{\l'm'}
	\,. \label{Eq:uest:tau}
}
The filtered multipoles are given by $\oT_{\l m}=Q_\l\{\bR{C}^{-1}\hT\}_{\l m}$ where $\bR{C}$ is the covariance of the observed temperature anisotropies and $Q_\l$ is the quality factor. 
We follow \cite{P15:phi} to compute $\T'_{\l m}\equiv \{\bR{C}^{-1}\hT\}_{\l m}$ in which the signal covariance is diagonal in harmonic space and the noise covariance is diagonal in pixel space with a white noise level of $30\,\mu$K-arcmin for unmasked pixels and infinite noise for masked pixels. 
Denoting $N_\l$ as a noise power spectrum of the $30\,\mu$K-arcmin white noise divided by the square of the beam function, we choose the quality factor as $Q_\l=1/(\CTT_\l+N_\l)/C_\l^{\T'\T'}$ \cite{P15:phi}. 
\footnote{For details, see Appendix A.1 of \cite{P15:phi}.}. 
The temperature fluctuations at $100\leq \l\leq 2048$ are used for reconstruction \cite{P15:phi}. 
The mean-field bias, $\ave{\uestt_{LM}}$, is nonzero due to e.g. Galactic and point source masks, and inhomogeneous instrumental noise. We compute it by averaging over simulation realizations. 
The unbiased estimator is then given by:
\al{
	\estt_{LM} = A^\tau_L(\uestt_{LM}-\ave{\uestt_{LM}}) 
	\,. \label{Eq:estt}
}
Here, the estimator normalization $A^\tau_L$ is defined as:
\al{
	A^{\dtau}_L &= \left[\frac{1}{2L+1}\sum_{\l\l'} \frac{1}{2}F_\l F_{\l'} (f^{\dtau}_{\l L\l'})^2 \right]^{-1}
	\,, \label{Eq:norm}
}
with $F_\l\equiv Q_\l /(\CTT_\l+N_\l)$.
Using a non-zero $\dtau_{LM}$ simulation, we checked that the cross-spectrum between reconstructed and input $\dtau_{LM}$ agrees with the input $\dtau$ power spectrum within $\alt 3\%$; we therefore do not further correct the estimator normalization in our analysis. 

The estimator described in \eqs{Eq:uest:tau,Eq:estt}, however, suffers from substantial contamination by other types of mode couplings in temperature, namely those induced by lensing and point sources. 
To mitigate these biases, we adopt a ``bias-hardened'' estimator for the optical-depth estimator, $\estt^{\rm BH}_{LM}$ \cite{Namikawa:2012:bhe}. The estimator is expressed as a linear combination of three estimators; the optical depth, $\estt_{LM}$, lensing potential, $\estp_{LM}$, and square of point-source fluctuations, $\ests_{LM}$, so that $\estt^{\rm BH}_{LM}$ is immune to the bias from lensing and point-source fields (see \cite{HuOkamoto:2001,OkamotoHu:quad,Namikawa:2012:bhe,Osborne:2013nna} for the details of $\estp_{LM}$ and $\ests_{LM}$). The bias-hardened estimator is given by \cite{Namikawa:2013:bhepol}: 
\al{
    \estt^{\rm BH}_{LM} 
    = \sum_{x=\dtau,\phi,\src} \{\bR{R}^{-1}_L\}^{\dtau,x} \widehat{x}_{LM}
    \,. \label{Eq:bhe}
}
The elements of the above matrix are defined as:
\al{
    \{\bR{R}_L\}^{a,b} = \frac{A_L^a}{2L+1}\sum_{\l\l'} 
    \frac{1}{2} F_\l F_{\l'} f^a_{\l L\l'} f^b_{\l L\l'}
    \,, \label{Eq:response}
}
where $a,b$ are either $\dtau$, $\grad$ or $\src$, and $f^\grad_{\l L\l'}$ and $f^\src_{\l L\l'}$ are the functions characterising the mode mixing in temperature anisotropies by lensing and point-source fields, respectively \cite{OkamotoHu:quad,Osborne:2013nna}. 
The estimator normalization, $A^{\grad}_L$ and $A^{\src}_L$, are given by \eqs{Eq:norm} but replacing $f^{\dtau}_{\l L\l'}$ with $f^{\grad}_{\l L\l'}$ and $f^{\src}_{\l L\l'}$, respectively. 
A bias-hardened estimator which only mitigates the lensing contribution 
can be obtained by setting $\{\bR{R}_L\}^{\dtau,\sigma}=\{\bR{R}_L\}^{\grad,\sigma}=\{\bR{R}_L\}^{\sigma,\dtau}=\{\bR{R}_L\}^{\sigma,\grad}=0$ 
and $\ests_{LM}=0$ in \eq{Eq:bhe}. 
The bias-hardened estimator minimizes any contamination by lensing and point sources; minimizing lensing contamination is especially critical because the correlation between the $y$-map and lensing is much larger than $\Cty$. 
Note that at the resolution of Planck, the pressure profile of all but the largest and most nearby clusters is not resolved, so that they appear as point sources and can be accounted for by this procedure. 
The bias-hardened estimator has a larger reconstruction noise than the non bias-hardened estimator. 
In our case, using the bias-hardened estimator increases the reconstruction noise 
by $\sim 0-40\%$ depending on scale. 

\section{Data and Simulation} \label{sec:data}

We use CMB data from the Planck 2015 release in our analysis \footnote{\url{http://pla.esac.esa.int/pla/\#home}}. 
The procedure for measuring $\dtau_{LM}$ is similar to that described in \cite{Namikawa:2017:plktau} and the Planck 2015 lensing paper \cite{P15:phi}.
We use temperature anisotropies from the Planck component separated maps (\SMICA\ or \NILC) \footnote{The Planck polarization data is noisy and the noise level of the reconstructed $\dtau$ using polarization is at least $1$-$2$ orders of magnitude larger than that of temperature. Therefore, the use of Planck polarization data does not help much for the reconstruction of $\dtau$.} in order to reconstruct $\dtau_{LM}$. 
Following \cite{P15:phi}, we use temperature multipoles between $\l=100-2048$ in our analysis. We apply the lensing mask, $W^{\rm CMB}(\hatn)$, used in the lensing trispectrum analysis to avoid Galactic foreground and point sources, and make use of the simulations of CMB signal and noise realization publicly available on 
\NERSC\ \footnote{\url{https://crd.lbl.gov/departments/computational-science/c3/c3-research/cosmic-microwave-background/cmb-data-at-nersc/}}. 

In addition, we use the Planck \NILC\ or \MILCA\ $y$-map for our cross-correlation measurement. 
In our baseline analysis, we use the full \NILC\ $y$-map and apply the mask obtained by combining the Galactic mask that retains $60$\% of the sky and point-source mask provided by Planck \cite{P16:tSZ}. 
To verify the stability of our results, we test the dependence of our analysis on the component separation method 
(\NILC/\MILCA) and the masking of the $y$-map. 
To evaluate the errors of the measured cross spectrum, we generate Gaussian random field simulations of the $y$-map whose power spectrum is consistent with the measured spectrum. The expected correlation with the $\dtau$ map is sufficiently small that we may safely neglect it in our simulations. 

In our cross-spectrum measurement, we multiply the $y$-map by an apodized mask, 
$W^{\rm y}(\hatn)$, calculate the spherical harmonic coefficients, $y_{LM}$, and deconvolve them with a $10$ arcmin Gaussian beam. 
We finally cross-correlate $\estt^{\rm BH}_{LM}$ and $y_{LM}$ to obtain 
$\Cty$. The cross-spectrum is divided by $\int {\rm d}^2\hatn [W^{\rm CMB}(\hatn)]^2W^{\rm y}(\hatn)$ to correct the normalization for masking effects. We also use the auto power spectrum of the optical depth anisotropies to constrain reionization. The methodology for measuring $\Ctt$ from the reconstructed optical depth anisotropies is the same as that of \cite{Namikawa:2017:plktau}.

\section{Results}

\begin{figure}[t]
\centering
\includegraphics[width=80mm,clip]{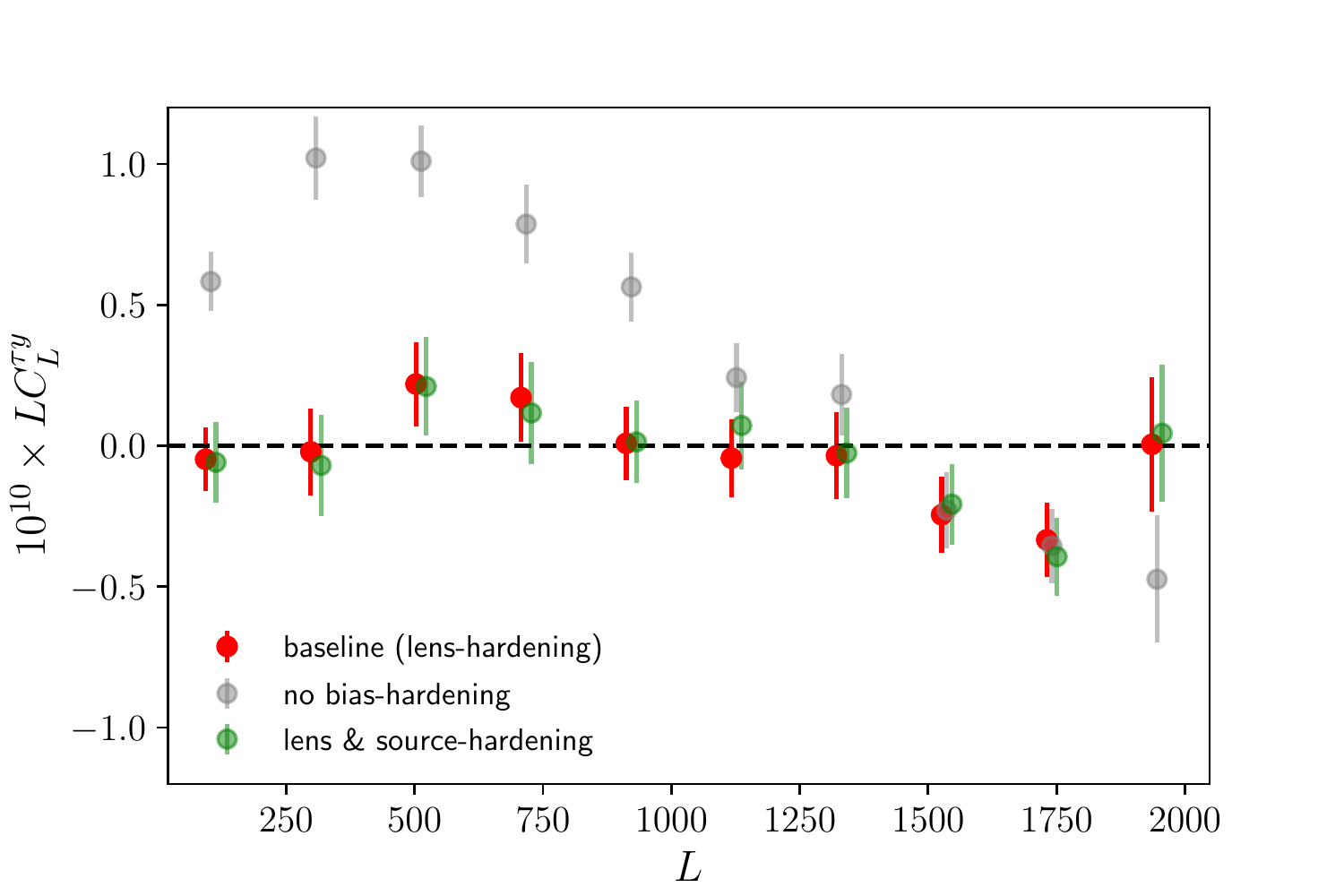}
\caption{
Optical depth -- Compton-$y$ cross-spectrum measurements with different estimators of the CMB optical depth fluctuations $\dtau$; the bias-hardened estimator to mitigate lensing bias (red, baseline) and the bias-hardened estimator to mitigate both lensing and point source bias (green, lens \& source hardening). A signal is not significantly detected but this measurement can be used to place constraints on reionization parameters. Note that, to illustrate the importance of bias-hardening, we also plot the result of naively using a simple quadratic estimator for the optical depth (grey, no bias-hardening). 
}
\label{fig:bhe}
\end{figure}

Fig.~\ref{fig:bhe} contains one of the main results of our analysis: the measurement of $\Cty$. 
We show three different cross-spectra in this plot, which differ in how they reconstruct $\dtau$: $\dtau$ is reconstructed either with the standard quadratic estimator, with a bias-hardened estimator to mitigate lensing, or with a bias-hardened estimator that mitigates both lensing and point-source contributions. 
It can be seen that the lensing contamination of $\estt_{LM}$ produces a significant bias in the cross spectrum measurement since the lensing and $y$-map are well correlated \cite{Hill:2013dxa}. 
There is almost no difference in the cross spectrum from the lensing-hardened and lensing-source-hardened estimators, indicating that the point-source-like extragalactic foreground contribution (e.g., from tSZ and CIB leaking through the $\dtau$ estimator and correlating with the $y$-map) is negligible in the cross-spectrum measurement. We compute the chi-squared probability-to-exceed ($\chi^2$-PTE) for the baseline spectrum with respect to null, finding that the value is $0.18$ and the spectrum is consistent with null.

\begin{figure}[t]
\centering
\includegraphics[width=80mm,clip]{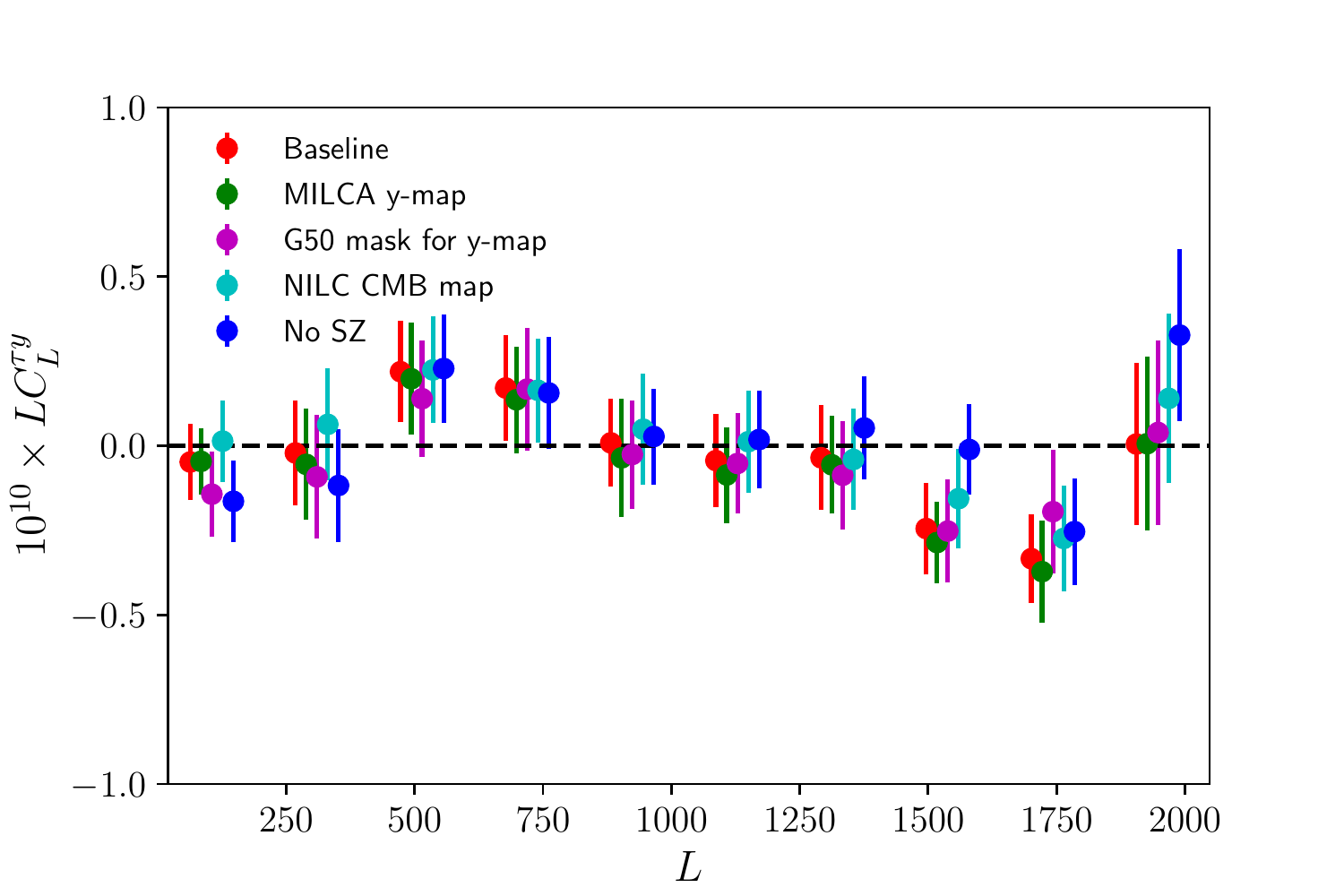}
\caption{
Cross-spectrum measurements as in Fig. ~\ref{fig:bhe} but with variations of the standard analysis choices, i.e., the use of a MILCA $y$-map, an aggressive Galactic mask (G50 mask) for the $y$-map, a NILC component separated CMB map, or an SZ-deprojected (No SZ) CMB map. The stability of the result to variations in these analysis choices suggests that any contamination of the measurement by galactic and extragalactic foregrounds is small. 
}
\label{fig:compsep}
\end{figure}

Fig.~\ref{fig:compsep} shows the measured cross spectra for different analysis choices. Using the different component separation methods for the $y$-map and CMB temperature anisotropies, we can test for contamination by extragalactic foregrounds such as tSZ, CIB and point sources. Using the different Galactic masks, we can also test for Galactic foreground contaminants. 
Overall, we only find small shifts to the bandpowers obtained when we vary the analysis choices. We compute the $\chi^2$-PTE for each case, finding that the values are in the range between $0.21$ and $0.74$ and that all these measured cross spectra are consistent with null within $2\,\sigma$.  

\section{Data interpretation}

We assume that both reionization at high redshift and clusters at low redshift contribute to the $\Cty$ signal. The total signal can be expressed as:
\al{
    \Cty = C^{\tau y,{\rm re}}_L 
    + \alpha \times C^{\tau y, {\rm lowz}}_L 
    \,,
}
where the first term is equivalent to \eq{eq:cltauy} and $C^{\tau y, \rm lowz}_L$ is the contribution from clusters calculated from cosmological simulations \cite{Battaglia2010} and diffuse sources and $\alpha$ is a free parameter which takes into account the uncertainties in the estimation. 

We perform Markov chain Monte Carlo (MCMC) analysis using \textit{emcee} software package \cite{emcee} to constrain the reionization parameters such as the average temperature of ionized bubbles $T_{\rm b}$, characteristic radius of ionized bubbles $R_{\rm b}$, and their distribution $\sigma_{\rm lnr}$ \cite{Wang:2006}. We also marginalize over the normalization factor of the low-$z$ contribution, $\alpha$. Our model of $\Cty$ depends on these parameters as well as on the reionization history and the bias of ionizing sources, $b$. For simplicity, we keep $b=6$ fixed and check the robustness of our results against $b$ below. 
We place the following flat logarithmic priors on $0.01$\,Mpc $<R_{\rm b}<10$\,Mpc, $0.01<\sigma_{\rm lnr}<1.5$, $4<\log(T_{\rm b}) <7$, and $0.1<\alpha<10$. 

\begin{figure}[t]
\includegraphics[width=70mm,clip]{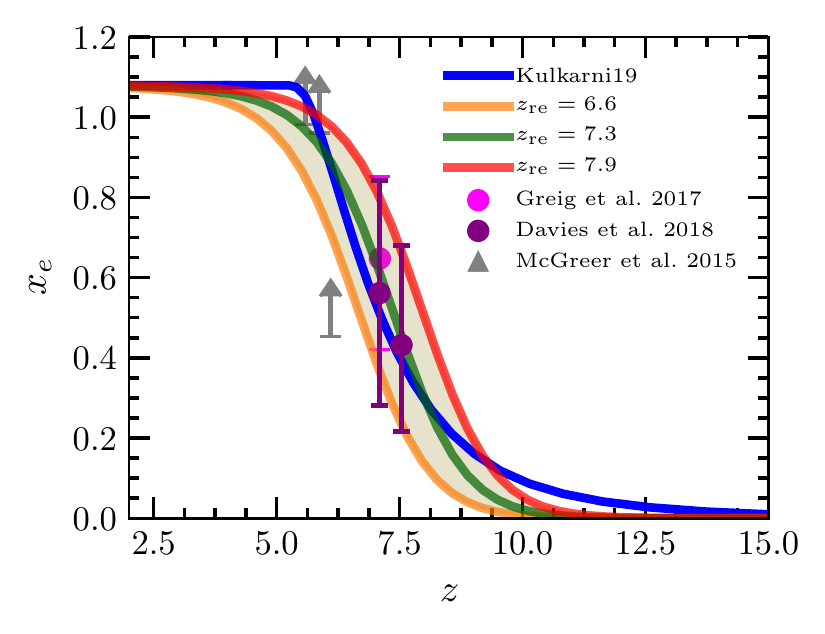}
\caption{
Evolution of the ionization fraction of the Universe inferred from the dynamic range radiative transfer simulation \cite{Kulkarni:2018}. For comparison, we show the \textit{tanh} reionization model varying the redshift of reionization, $z_{\rm re}$. Error bars are shown from the study of the spectral shape of the two highest-redshift Quasars \cite{Davies:2018,Greig:2016} and Ly$\alpha$ and Ly$\beta$ forests \cite{McGreer:2014}.
}
\label{fig:reion}
\end{figure}

We also need a model describing the evolution of the ionization fraction. In Fig.\,\ref{fig:reion}, we show the reionization history from the Sherwood simulation \cite{Kulkarni:2018} which gives rise to $\taumean=0.054$. For comparison we also show \textit{tanh} models of ionization histories parametrized by the redshift of reionization $z_{\rm re}$ and the duration of reionization $\Delta z$ as $x_{\rm e}(z)=(f/2)[1+\tanh{((y_{\rm re}-y)/\Delta y_{\rm re}})]$, where $y(z)=(1+z)^{1.5}$, $y_{\rm re}=y(z_{\rm re})$ and $\Delta y_{\rm re}=\sqrt{1+z_{\rm re}}\Delta z$ and $f=1.08$ for the first ionization of helium atoms. For a fixed $\Delta z=1.5$, we varied the redshift of reionization $z_{\rm re}$ to $6.6$, $7.3$ and $7.9$; the isotropic CMB optical depths that correspond to these reionization histories are $\taumean=0.047, 0.054$ and $0.061$ respectively. Unless mentioned specifically, we have used the Kulkarni19 \cite{Kulkarni:2018} reionization history for $\Cty$ calculation from semianalytic model.

\begin{figure}[t]
\includegraphics[width=70mm]{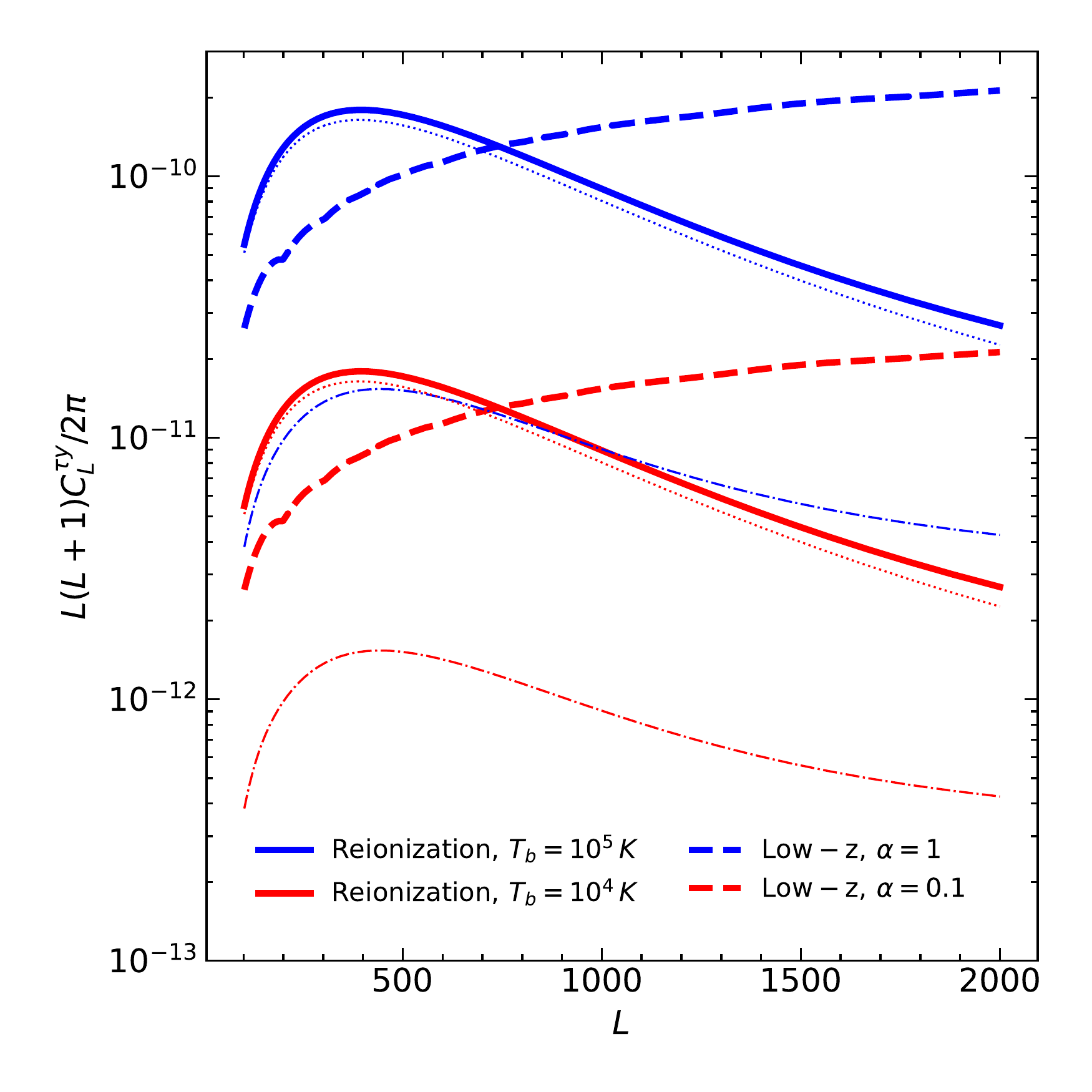}
\caption{The cross-correlation signal between $\dtau$ and $y$ arising from reionization (blue) and clusters (red). We assume a characteristic ionized bubble radius $R_{\rm b}=10$\,Mpc and distribution width $\sigma_{\rm lnr}=\ln(2)$, and we vary the ionization temperature $T_{\rm b}$ and $\alpha$ (a parameter rescaling the low-redshift cluster contribution to the cross-spectrum). Dashed dotted and dotted lines represent the one-bubble and two-bubble contributions. The cluster contribution (Low-z) is computed from \cite{Battaglia:2016}.}
\label{fig:cl_comparison}
\end{figure}

In Fig.\,\ref{fig:cl_comparison}, we compare the $\Cty$ signal from the reionization and cluster contributions. We use $R_{\rm b}=10$ Mpc and $\sigma_{\rm lnr}=\ln(2)$ for this $\Cty$ calculation. The cluster contribution in $\Cty$ is estimated using cosmological simulations \cite{Battaglia2010}. Here, $\dtau$ and $y$ maps were made at corresponding redshift slices and cross-correlated. The cross power spectra were then combined following the procedure in \cite{Battaglia2012b}. The reionization contribution becomes larger at large scales than that of cluster whereas the cluster contribution becomes dominant on small scales. For fixed $R_{\rm b}$ and $\sigma_{\rm lnr}$, the ionized bubble temperature scales the amplitude of the signal linearly keeping the shape of the spectra unchanged. $\alpha=0.1$ means that the actual cluster contribution in $\Cty$ is ten times smaller than that estimated from these simulations. 

\begin{figure}[t]
\centering
\includegraphics[width=85mm,clip]{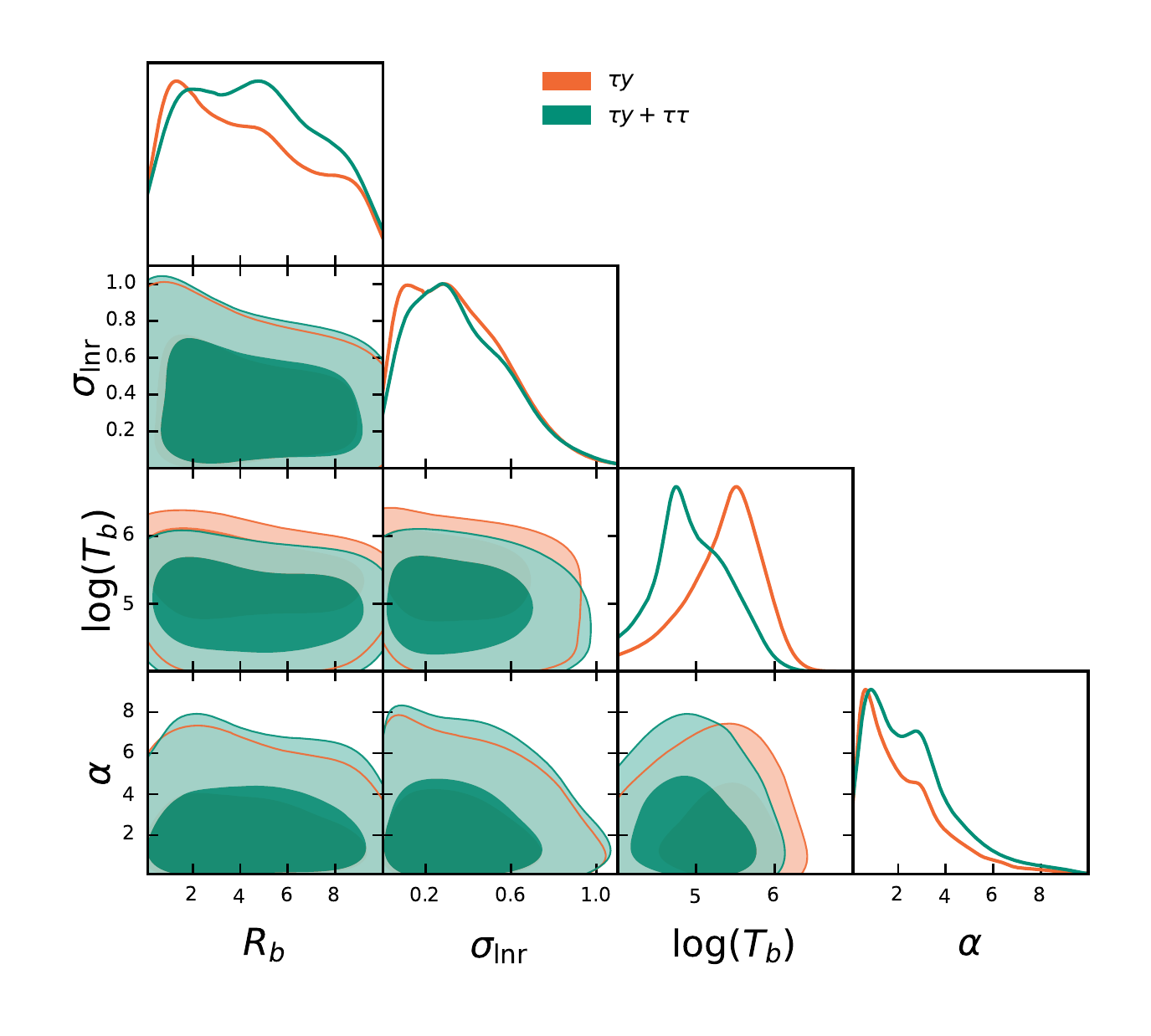}
\caption{
$1\,\sigma$ and $2\,\sigma$ constraints on reionization parameters arising from our Planck measurements of $\Cty$ shown in Fig. ~\ref{fig:bhe} (as well as $\Ctt$). We show constraints on the temperature $T_{\rm b}$ and radius $R_{\rm b}$ of ionized bubbles and their distribution along with the normalization factor of the cluster contribution $\alpha$.
}
\label{fig:contour_joint_measurement}
\end{figure}

\begin{figure}[t]
\centering
\includegraphics[width=85mm,clip]{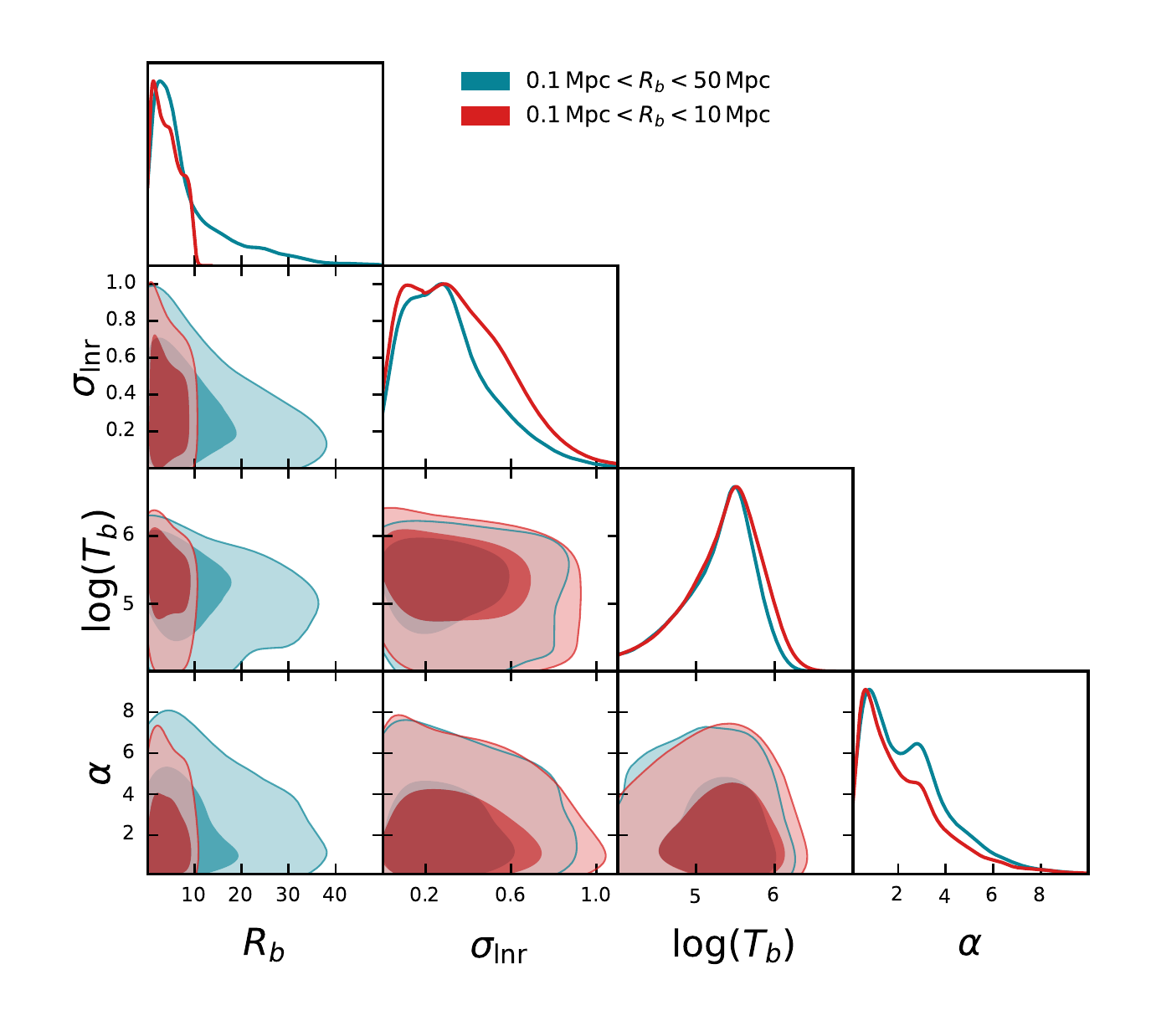}
\caption{
Constraints as in Fig.~\ref{fig:contour_joint_measurement} ($\Cty$-derived constraints), but we explore the sensitivity to priors on the characteristic bubble radius $R_{\rm b}$ by imposing two different flat logarithmic priors on $0.01$\,Mpc $<R_{\rm b}<10$\,Mpc and $0.01$\,Mpc $<R_{\rm b}<50$\,Mpc. Most parameters appear insensitive to the details of this prior choice.
}
\label{fig:contour_r_prior}
\end{figure}

In Fig.\,\ref{fig:contour_joint_measurement}, we show the constraints on $R_{\rm b}$ and $\sigma_{\rm lnr}$ through a triangle plot showing $1\,\sigma$ and $2\,\sigma$ contours. We performed the analysis by fitting our reionization model with $\Cty$ alone and also for a joint analysis of $\Cty$ and $\Ctt$. 
We quote the $2\,\sigma$ upper limit on these parameters in Table \ref{tab:upper_limit}. 
We impose a hard prior $0.01\,{\rm Mpc}<R_{\rm b}<10\,$Mpc since the simulation studies suggest $R_{\rm b}\alt 10$\,Mpc.
Although the $2\,\sigma$ upper bound on $R_{\rm b}$ does not significantly change for the joint analysis with this prior, the temperature bound decreases by $45$\% for the joint analysis relative to the $\Cty$ only analysis. To check the validity of our results we also assume a flat prior on  $0.01\,{\rm Mpc}<R_{\rm b}<50\,$Mpc. We show dependencies of our result on the choice of priors on $R_{\rm b}$ in Fig.\,\ref{fig:contour_r_prior}. The constraint on $T_{\rm b}$ becomes $\sim 10\%$ tighter if we assume a prior $0.01\,{\rm Mpc}<R_{\rm b}<50\,$Mpc. However, the upper limit on the temperature of the ionized bubbles is more than one order of magnitude larger than the measurement from the Ly$\alpha$ forest \cite{2010MNRAS.406..612B}.

We quote the $2\,\sigma$ upper limit on these parameters in Table \ref{tab:upper_limit}. 
The upper limits on $R_{\rm b}$ and $T_{\rm b}$ increase by $\sim30\,\%$ and $\sim10\,\%$, respectively, if $z_{\rm re}$ varies from $7.9$ to $6.6$. Both the size and temperature of the ionized bubbles increases for the earlier reionization history due to higher density contrast responsible for the inhomogeneous reionization. We did not find any significant changes in the distribution of ionized bubbles for a delayed reionization process.

To check the robustness of our conclusions, we vary the priors on the bias to explore the $R_{\rm b}$-$\sigma_{\rm lnr}$ relation. For different choices of Gaussian prior on $b$ we did not find any significant changes on the $2\,\sigma$ upper limit on $R_{\rm b}$ and $T_{\rm b}$. Similarly, we use different reionization histories as shown in Fig.\,\ref{fig:reion} and find that the the $2\,\sigma$ bound on $R_{\rm b}$ and $T_{\rm b}$ changes by, at most, 10\% and 2\% respectively. 

\begin{table}[t]
\begin{tabular}{cccc}
\hline\hline
Priors on $R_{\rm b}$& \multicolumn{2}{c}{$[0.01,10]$} & \multicolumn{1}{c}{$[0.01,50]$} \\
\cline{1-4} 
Parameters & $\Cty$ & $\Cty$+$\Ctt$ & $\Cty$ \\
\hline
$R_{\rm b}$ & $9.4$ & $9.5$ & $23.1$ \\
$\sigma_{\rm lnr}$ & $0.81$ & $0.83$ & $0.74$ \\
$\log(T_{\rm b})$ & $6.11$ & $5.85$ & $6.05$ \\
$\alpha$ & $5.8$ & $6.3$ & $6.16$ \\
\hline
\end{tabular}
\caption{
We quote the $2\,\sigma$ upper bound on parameters describing reionization derived (in part) from our baseline $\Cty$ measurement from Planck data shown in Fig.~\ref{fig:bhe}.
}
\label{tab:upper_limit}
\end{table}

\begin{table}[t]
\begin{tabular}{ccccc}
\hline
CMB experiment & $f_{\rm sky}$ & $\theta$ [arcmin] & $\sigma_{\rm P}$\,[$\mu$K-arcmin] & $\alpha_{\rm del}$ \\
\hline
S4-Wide & $0.4$ & $2.0$ & $1.4$ & $0.2$ \\
S4-Deep & $0.04$ & $2.0$ & $0.42$ & $0.07$ \\
HD & $0.5$ & $0.2$ & $0.7$ & $0.1$ \\
\hline
\end{tabular}
\caption{
Our assumptions for the CMB experiments used to reconstruct $\delta\tau$. $f_{\rm sky}$, $\theta$, $\sigma_P$ and $\alpha_{\rm del}$ denote the observed sky fraction, angular resolution, white noise level in the polarization map, and residual fraction of lensing $B$-mode spectrum after delensing, respectively.
}
\label{table:cmbexp}
\end{table}

In our analysis, we consider only the temperature increase of the IGM due to the ionization of hydrogen and singly ionized helium. It should be noted that the IGM temperature is also heated up by $\sim 5000-10000$\,K due to the HeII reionization at $z>2.5-3$ \cite{McQuinn:2008am}. Including this effect would further tighten our constraints on the IGM temperature.  We also do not include the contribution of the cooling of different metal lines present in the IGM, and the cooling due to the adiabatic expansion and inverse Compton scattering off of the CMB. The temperature of the IGM depends on the shape of the radiation spectrum of ionizing sources; one part of it ionizes the neutral hydrogen atoms and rest of it increases the kinetic energy of the electron gas. In reality, this process is very complex as the shape of the spectral energy distribution varies from one source to another source. It also involves the modeling of adiabatic cooling due to expansion, the temperature fluctuation in the IGM due to the inhomogeneous nature of reionization, and the influence of low mass galaxies in the reionization process. 

\section{Forecasts}

In this section we investigate how the constraints on reionization parameters, such as the temperature and ionized bubble size, will improve for future high-resolution and low-noise CMB experiments.

In our forecasts, we define a fiducial model for $\Cty$ and $\Ctt$ with the parameters $R_{\rm b}=5$\,Mpc, $\sigma_{\rm lnr}=\ln(2)$, $T_{\rm b}=5\times10^4$\,K, and $\alpha=1$. For the $y$-map, we consider the Planck and the Probe of Inflation and Cosmic Origins (PICO) \cite{Hanany:2019lle} cases. We use the observed $C_L^{yy}$ for the Planck case. In the PICO case, we assume a noiseless $y$-map spectrum since the PICO $y$-map is signal dominated at the scales of our interest \cite{Hanany:2019lle}. For reconstructing $\dtau$, we consider the CMB experiments summarized in Table~\ref{table:cmbexp}; ``S4-Wide'' corresponds to the CMB-S4 wide area legacy survey. ``S4-Deep'' denotes the CMB-S4 delensing survey, i.e. a sky fraction of few percent but with very deep observations to effectively perform delensing. The ``HD'' survey is similar to CMB-HD \cite{Sehgal:2019ewc}. We do not consider the PICO $\dtau$ since the expected noise level is higher than S4 due to the limited angular resolution. For simplicity, we adopt approximate noise levels and angular resolutions which mimic the above experiments. For the purpose of forecasts, we simplify the covariance of the $\Ctt$ and $\Cty$ spectra as being described by a diagonal matrix, i.e.: 
\al{
    \sigma^2(C^{\tau y}_b,C^{\tau y}_{b'}) = \delta_{bb'}\left[\sum_{L\in b}\frac{f_{\rm sky}(2L+1)}{A^{\dtau}_L C^{yy,\rm obs}_L} \right]^{-1}
    \,, \label{Eq:error:tauxy} \\
    \sigma^2(C^{\tau\tau}_b,C^{\tau\tau}_{b'}) = \delta_{bb'}\left[\sum_{L\in b}\frac{f_{\rm sky}(2L+1)}{2(A^{\dtau}_L)^2}\right]^{-1} 
    \,. 
}
We ignore $\Cty$ and $\Ctt$ in $\sigma^2(C^{\tau y}_b,C^{\tau y}_{b'})$ because $\Cty\alt 0.01\times(A^{\dtau}_LC^{yy}_L)^{1/2}$ and $\Ctt\alt 0.1\times A^{\dtau}_L$, respectively, for our fiducial model. The cross-covariance between $\Ctt$ and $\Cty$ is set to zero. The reconstruction noise is computed from the $EB$-estimator with CMB multipoles at $100\leq\l\leq4096$. To consider the removal of lensing contribution in $B$-modes (delensing), the lensing $B$-mode spectrum is scaled by $\alpha_{\rm del}$ in Table~\ref{table:cmbexp} whose values are determined by performing the iterative delensing procedure proposed by \cite{Smith:2010gu}. In this calculation, we ignore the fact that in practice we need a joint iterative estimate of $\dtau$ and lensing from a given CMB map \cite{Guzman:2021}. Although the joint estimate could enhance the noise level of $\dtau$, the impact is expected to be small; the two effects, optical-depth anisotropies and lensing, on CMB anisotropies are very different and the degradation of the noise level when bias-hardening against lensing is typically $10\%$ in the case of the quadratic estimator \cite{Namikawa:2012:bhe}. A thorough analysis of this joint estimate using optimal reconstruction methods is beyond the scope of our paper and will be addressed elsewhere. 

\begin{figure}[t]
\includegraphics[width=85mm,clip]{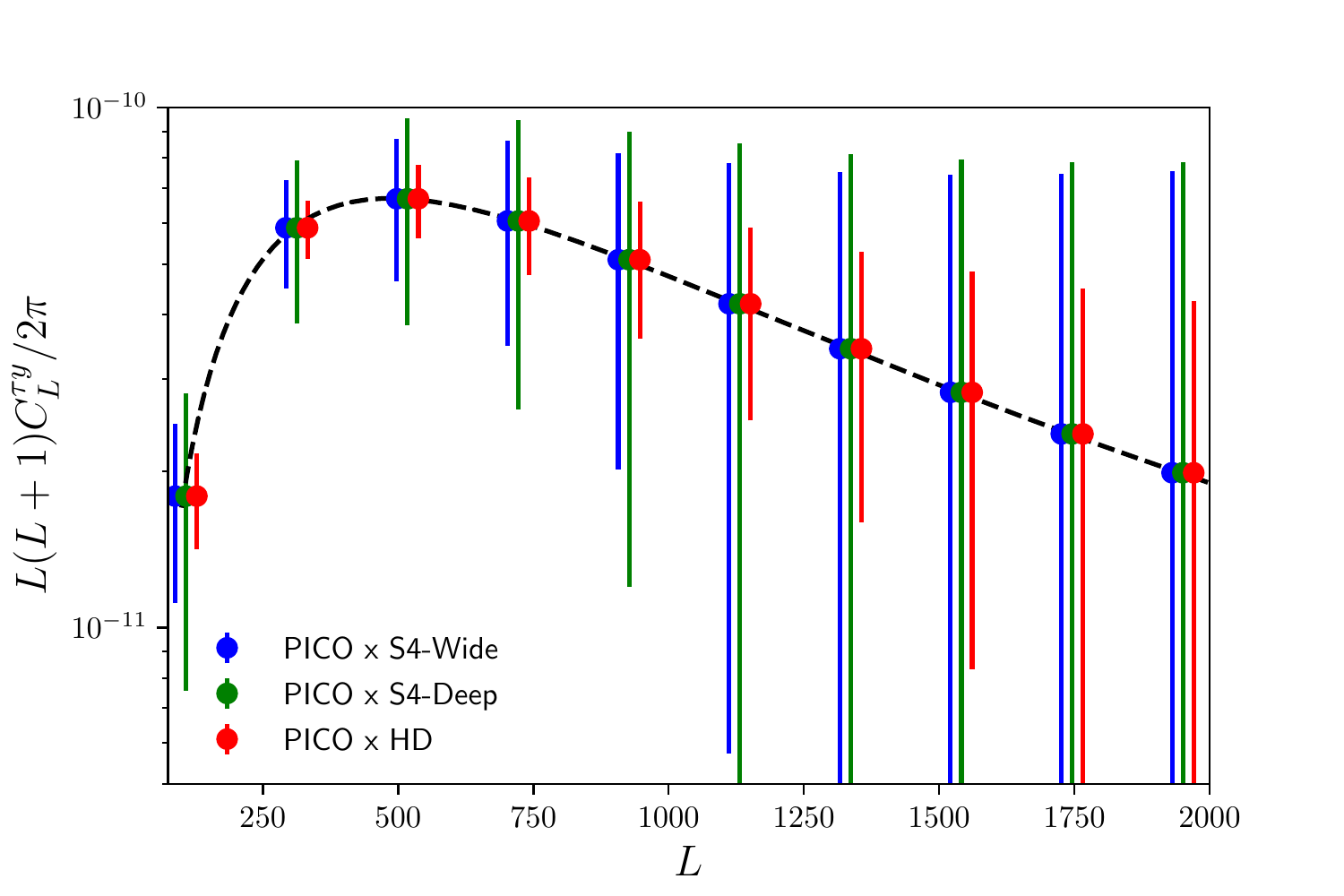}
\caption{Forecast cross-power spectrum between optical-depth fluctuations and a $y$-map, $\Cty$, with expected error bars shown for some of the experimental cases we consider. 
We assume $R_{\rm b}=5$\,Mpc, $\sigma_{\rm lnr}=\ln(2)$, and $T_{\rm b}=5\times10^4$\,K. It can be seen that for future CMB experiments, our forecasts indicate that significant detections of $\Cty$ appear possible (assuming our fiducial reionization model). 
}
\label{fig:forecast-tyspec}
\end{figure}

\begin{figure}[t]
\includegraphics[width=85mm]{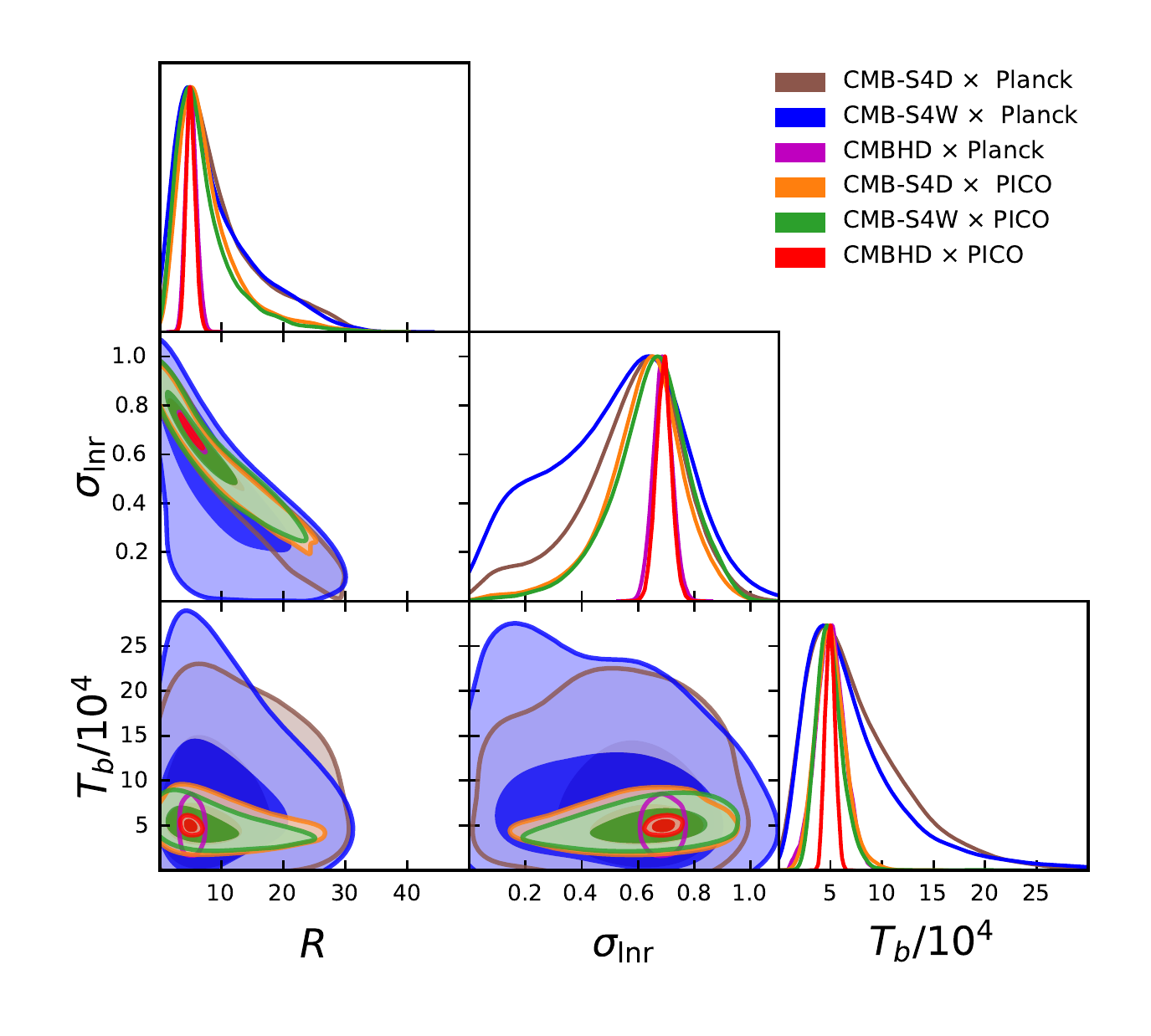}
\caption{
Forecast of reionization parameter constraints expected from future experiments (the experiment named first is the experiment providing the $\dtau$ map, the experiment named second provides the $y$-map). We show 68\% and 95\% contours derived from a joint analysis of $\Cty$ and $\Ctt$. It can be seen that precise constraints appear possible, especially when a signal-dominated PICO-like $y$ map is available. 
}
\label{fig:forecast}
\end{figure}

Fig.~\ref{fig:forecast-tyspec} shows the cross-power spectrum, $\Cty$, with the expected error bars computed from \eq{Eq:error:tauxy}. Future experiments have much better sensitivity to detect the cross spectrum due to both improving the reconstruction noise level of $\dtau$ with lower polarization map noise and the $y$-map with better component separation from multiple frequency data. Combining $\dtau$ from CMB-S4 and a $y$-map from PICO, and assuming our fiducial model, the signal-to-noise of $\Cty$ becomes $\simeq 7$. Note that the signal-to-noise linearly scales with $T_{\rm b}$. Using $\dtau$ from CMB-HD and $y$-map from PICO further improves the signal-to-noise of $\Cty$ to be $\simeq 13$. In Fig.\,\ref{fig:forecast} we forecast the constraints on the reionization model parameters for future CMB experiments. In this analysis, we assume a Gaussian prior on $\alpha=1 \pm 0.1$. Current measurements of density and pressure profiles from kSZ and tSZ cross-correlations, respectively, with luminous red galaxies are already observed at $\sim 8\,\sigma$ or better \cite{Schaan2020,Amodeo2020}, so it reasonable to assume that future measurements will improve our understanding of the low-z contribution to $\Cty$ at the 10\% level. In Table\,\ref{tab:forecast} we quote the reionization parameter constraint forecasts. We find that for current $y$ maps from Planck, the constraints on reionization parameters are fairly weak, unless a futuristic $\dtau$ measurement from CMB-HD is provided. However, if a signal-dominated $y$-map, e.g. from PICO, is available, combining the optical depth anisotropies measured from CMB-S4 could tightly constrain reionization parameters, allowing a detection of the temperature of $5\times10^4\,$K at more than $3\,\sigma$. 

Note that we have ignored any possible contamination in the cross-spectrum measurements in the forecast. For example, as we discussed above, leakage from CIB and kSZ propagating through the quadratic $\dtau$ estimator could bias the cross-spectrum. Although this bias should be negligible in our measurement with current data, in future CMB experiments, we will use CMB anisotropies up to very small scales, which could lead to a non-negligible bias in the measurement. Point-source hardening can capture some amount of this bias, but it is not clear how well point-source hardening will work for higher-precision experiments. However, even if simple point-source hardening does not completely subtract the bias from CIB and kSZ, we can construct a bias-hardened estimator with source profiles as in \cite{Sailer:2020:bhe}, or employ more sophisticated multifrequency mitigation techniques. Further, in future experiments, the signal-to-noise of $\dtau$ is almost determined by precision of polarization data. The polarization-based reconstruction of $\dtau$ should be much less affected by the above contaminants. Therefore, astrophysical contaminants are not expected to significantly bias $\Cty$ measurements in future experiments.

\section{Discussion}

The temperature measurement of the IGM carries important information about the source of the reionization. Various ionizing sources emit photons with different energies which later heat up the surrounding neutral medium by the propagation of ionization front. The temperature of the IGM depends on the velocity of the ionization front and the energy of the ionizing photons that determines the volume average ionization fraction at a particular redshift \cite{Lidz:2014jxa}. Radiative transfer simulations show that temperature of IGM can reach $25000$\,K$-\,30000$\,K during the end of reionization if the velocity of ionization front reaches to $10^4$\,km/s \cite{DAloisio:2018rzi,Mcquinn2012, Finlator:2018lbl}. 

Observation of the Ly$\alpha$ forest around Quasar proximity zones is an alternative and powerful probe of the IGM temperature. However, it does not necessarily represent the global properties of the IGM temperature at high redshift as there are very few Quasars at $z>6$ and their environments are atypical. The temperature is determined using radiative transfer simulations with the observed Ly$\alpha$ forest power spectrum, which means that the temperature is, to some extent, model dependent. Recent measurements of temperature from the high resolution Ly$\alpha$ forest spectra show that the temperature is $\sim 12000\pm 2000$\,K \cite{Gaikwad:2020art} at $z\sim 5.8$. The first constraint of the IGM temperature at $z=6$ is $23600^{+9200}_{-9300}$\,K \cite{Bolton:2010gr}. 

In the future, the precise measurement of the temperature fluctuations of individual HII bubbles will enrich our understanding about the metallicity of the ionizing stars. The temperature of HII regions ionized by the intermediate metallicity stars is two to three times higher than the HII bubbles ionized by the low metallicity stars \cite{Katz:2018xle}. On the other hand, to keep the gas ionized during the EoR, the virial temperature of the halos for star formation should be higher than $10^4$\,K. At $z>6$ high mass halos are rare but there are many low mass halos. It is possible that population III stars could have ionized the neutral medium during EoR as they are heavy and have higher effective temperatures. Studies of spectral lines show that effective temperature of Population III and II with mass 15\,$M_\odot$ are $\sim 63000$\,K and $36000$\,K \cite{Tumlinson:1999iu}. It could be the case that Population II stars contributed during the end of reionization. Such a high-temperature scenario would be tested by measuring $\Cty$. 

\begin{table}[t]
\begin{tabular}{cccc}
\hline\hline
Experiments  & $R_{\rm b}$ [Mpc]  & $\sigma_{\rm lnr}$ & $T_{\rm b}$ [K] \\
\hline
S4-Deep$\times$Planck & $9.5^{+2.4}_{-7.8}$ & $0.57^{+0.22}_{-0.13}$   &  $77000^{+23000}_{-60000}$ \\
S4-Wide$\times$Planck & $9.0^{+2.7}_{-7.9}$  & $0.51^{+0.29}_{-0.20}$ &  $80000^{+13000}_{-63000}$  \\
HD$\times$Planck & $5.19^{+0.77}_{-0.96}$ &  $0.688^{+0.034}_{-0.034}$  & $50800^{+14000}_{-13000} $  \\

S4-Deep$\times$PICO & $7.9^{+1.7}_{-5.7}$ & $0.62^{+0.15}_{-0.11}$  & $51000^{+12000}_{-16000}$  \\

S4-Wide$\times$PICO & $7.2^{+1.6}_{-5.4}$  & $0.64^{+0.15}_{-0.11}$  & $49700^{+9500}_{-15000}$ \\

HD$\times$PICO & $5.09^{+0.66}_{-0.79}$ & $0.691^{0.028}_{-0.032}$  & $49800^{+4500}_{-5100}$  \\
\hline
\end{tabular}
\caption{
$1\,\sigma$ uncertainties on the reionization parameters forecast for future CMB experiments. We here consider reconstructing the fluctuations in optical depth for configuration of the first experiment and then cross-correlating with the Compton-$y$ parameter for the second experiment. Note that the lower initial values of $T_{\rm b}$ of the fiducial model, the SNR changes approximately linearly.
}
\label{tab:forecast}
\end{table}

\section{Conclusions}

In this paper, we present the first constraint on the optical depth -- Compton-y cross-correlation $\Cty$ from Planck CMB data and use it to place bounds on several important properties of reionization: the temperature of ionized bubbles $T_{\rm b}$ as well as the bubbles' characteristic radius $R_{\rm b}$ and size distribution.

We also present forecasts of how well this new observable can constrain reionization parameters with future experiments. Our forecasts show that future experiments will be able to measure $\Cty$ and $\Ctt$ signals and thus reionization parameters such as $T_{\rm b}$ and $R_{\rm b}$ with much better sensitivity. In particular, for a fiducial reionization model, we predict that the CMB-HD -- PICO cross-correlation should enable a precise measurement of the reionization parameters: $T_{\rm b}\simeq 49800^{+4500}_{-5100}$\,K and $R_{\rm b}\simeq 5.09^{+0.66}_{-0.79}$\,Mpc; with a CMB-S4 -- PICO cross-correlation, informative constraints on reionization parameters (that are only $\sim 2$ times weaker than with CMB-HD) can also be achieved. Further improvements to these constraints are likely achievable if we can subtract off the limiting low-redshift contribution to the Compton-y map using our knowledge of the distribution and properties of gas at $z<2$; we defer a detailed discussion of this possibility to future work.

The unique aspect of the $\Cty$ signal is that it can be used as a direct probe of the reionization temperature. In a joint anaysis with $\Ctt$, the details of the reionization morphology do not affect the temperature constraint substantially since the model of the reionization morphology is constrained significantly by $\Ctt$.  

Future measurements of this $\Cty$ signal will hence complement other CMB constraints on reionization. Thus far, CMB data have constrained the duration and redshift of reionization via the sky-averaged optical depth \cite{Pagano:2019} and the upper limits on the kSZ power spectrum \cite{P16:reion,Reichardt:2020}. In the future, high resolution, low noise observations from ground-based, CMB experiments such as AdvACT \cite{Henderson:2015}, SPT-3G \cite{Benson:2014}, SO \cite{SimonsObservatory}, and CMB-S4 \cite{CMBS4} will provide new insights into reionization. Recent studies by \cite{Feng:2018:3pt,Feng:2018:tau-phi} propose using cross-correlation between $\dtau$ and large-scale structure tracers, showing that polarization data from futuristic experiments will be useful to probe reionization properties. In addition, LiteBIRD \cite{LiteBIRD} will achieve a cosmic-variance limited measurement of large-scale $E$ modes, providing a constraint on the isotropic part of the optical depth, $\taumean$, with $\sigma(\taumean)\simeq 2\times10^{-3}$. Looking ahead, combined analyses of $\Cty$, the kSZ power-spectrum and trispectum \cite{Alvarez:2020}, and the large-scale polarization can be expected to significantly improve our knowledge of reionization from CMB observations alone. These CMB observations will be independent of and complementary to new insights gained from 21cm experiments.

In addition to $\Cty$ on which we have focused, we can consider several other cosmological observables to be cross-correlating with $\dtau$ as a probe of reionization. The kSZ signal is one of the candidates among those observables. Note that kSZ contains modes along the line-of-sight while $\dtau$ contains modes perpendicular to the line-of-sight. Therefore, a naive cross spectrum between kSZ and $\dtau$ would be very small analogs to the cross spectrum between kSZ and CMB lensing \cite{Smith:2018:kSZ}. Instead, the cross spectrum between a square of kSZ signals and $\dtau$ does not suffer from the above issue. This statistic is a bispectrum of the fluctuations. For reionization, the fluctuations could be large since a reionization process would be highly non-Gaussian. To evaluate the bispectrum, however, we need an expensive simulation or a model to describe its non-Gaussianity. This calculation is involved and we leave this study for our future work.


\begin{acknowledgments}
We thank Simone Ferraro, Colin Hill, Vid Ir\v{s}i\v{c}, Girish Kulkarni, and Daan Meerburg for providing helpful and constructive comments on the draft. TN and BDS acknowledge support from an Isaac Newton Trust Early Career Grant and from the European Research Council (ERC) under the European Unions Horizon 2020 research and innovation programme (Grant agreement No. 851274). TN further acknowledges support from JSPS KAKENHI Grant Number JP20H05859 and World Premier International Research Center Initiative (WPI), MEXT, Japan. BDS further acknowledges support from an STFC Ernest Rutherford Fellowship. NB acknowledges support from NSF grant AST-1910021. For numerical calculations, we used resources of the National Energy Research Scientific Computing Center (NERSC), a U.S. Department of Energy Office of Science User Facility operated under Contract No. DE-AC02-05CH11231. The Flatiron Institute is supported by the Simons Foundation.
\end{acknowledgments}

\appendix
\section{Bias-hardened estimator for CMB optical depth}

Here, we derive the bias-hardened estimator which is used for the mitigation strategy of the lensing and point-source contributions in estimating $\dtau$.  

We first note that the lensing produces similar mode couplings in the temperature anisotropies. The quadratic estimator for the optical depth anisotropies given above is then the same as that for the CMB lensing potential, $\grad$, but with a different weight function. In the case of CMB lensing, the mode coupling of \eq{Eq:tau-mixing} is given by \cite{OkamotoHu:quad}:
\al{
	\ave{\hT_{\l m}\hT_{\l'm'}}\rom{CMB} 
		&= \sum_{LM}\Wjm{\l}{\l'}{L}{m}{m'}{M} f^\phi_{\l L \l'} \grad^*_{LM} 
	\,, \label{Eq:phi-mixing}
}
where $f^{\grad}_{\l L\l'}$ is defined as
\al{
	f^{\grad}_{\l L\l'} 
    &= -\gamma_{\l L\l'}\sqrt{L(L+1)}
    \bigg[\sqrt{\l'(\l'+1)}\Wjm{\l}{L}{\l'}{0}{1}{-1}\CTT_{\l'}
    \notag \\ 
    &\qquad +\sqrt{\l(\l+1)}\Wjm{\l'}{L}{\l}{0}{1}{-1}\CTT_{\l}\bigg]
	\,. \label{Eq:weight:phi}
}
The estimator for the CMB lensing potential is then obtained in the same form as \eq{Eq:uest:tau} but replacing $f^{\dtau}_{\l L\l'}$ with $f^{\grad}_{\l L\l'}$. The estimator normalization is also obtained in the same way. 

The mode mixing is also induced by point sources and we can estimate the point-source fields using the quadratic estimator. The temperature fluctuations have the following additive term in the presence of the point source, and more generally any additive fields, $s(\hatn)$, including inhomogeneous noise: 
\al{
	\hT(\hatn) &= \T(\hatn) + s(\hatn) 
	\,. 
}
Assuming $\ave{s(\hatn)s(\hatn')}\equiv\delta(\hatn-\hatn')\src(\hatn)$ \cite{Namikawa:2012:bhe}, 
the mode coupling between different temperature multipole has the following form: 
\al{
	\ave{\hT_{\l m}\hT_{\l'm'}}\rom{CMB} 
		&= \sum_{LM}\Wjm{\l}{\l'}{L}{m}{m'}{M} f^{\src}_{\l L \l'} \src^*_{LM} 
	\,, \label{Eq:src-mixing}
}
where $\src_{LM}$ is the harmonic coefficients of $\src(\hatn)$ 
and the weight function is defined as \cite{Osborne:2013nna,P15:phi}:
\al{
    f^{\src}_{\l L\l'} 
    &= \gamma_{\l L\l'}\Wjm{\l}{L}{\l'}{0}{0}{0}
    \,. \label{Eq:weight:src}
}

\eqs{Eq:phi-mixing,Eq:src-mixing} mean that the lensing and point source 
fields also contribute to the estimator of the optical depth anisotropies in 
\eq{Eq:uest:tau}. Using \eqs{Eq:tau-mixing,Eq:phi-mixing,Eq:src-mixing}, 
the estimator of the optical depth anisotropies has the following biases 
from the lensing and point source fields: 
\al{
    \ave{\estt_{LM}}_{\rm CMB} = \dtau_{LM}+R_L^{\dtau,\phi}\phi_{LM}+R_L^{\dtau,\src}\src_{LM} 
    \,. \label{Eq:estt-bias}
}
Here, the response function is given by \eq{Eq:response}. 

To mitigate the biases in the estimator of the optical depth anisotropies, 
let us consider biases to the lensing and point-source estimators, 
$\estp_{LM}$ and $\ests_{LM}$. 
The lensing and point source estimators are obtained by simply replacing the weight function $f^{\dtau}_{\l L\l'}$ with $f^{\grad}_{\l L\l'}$ or $f^{\src}_{\l L\l'}$ in \eq{Eq:uest:tau}. 
Using \eqs{Eq:tau-mixing,Eq:phi-mixing,Eq:src-mixing}, 
the lensing and point-source estimators are biased by: 
\al{
    \ave{\estp_{LM}}_{\rm CMB} &= \grad_{LM}+R_L^{\grad,\dtau}\dtau_{LM}+R_L^{\grad,\src}\src_{LM} 
    \,, \label{Eq:estp-bias} \\ 
    \ave{\ests_{LM}}_{\rm CMB} &= \src_{LM}+R_L^{\src,\grad}\grad_{LM}+R_L^{\src\dtau}\dtau_{LM} 
    \,. \label{Eq:ests-bias}
}
\eqs{Eq:estt-bias,Eq:estp-bias,Eq:ests-bias} are summarized as the following matrix form: 
\al{
    \begin{pmatrix} \ave{\estt_{LM}}_{\rm CMB} \\ \ave{\estp_{LM}}_{\rm CMB} \\ \ave{\ests_{LM}}_{\rm CMB} \end{pmatrix} 
    &= \begin{pmatrix} 
        1 & R_L^{\dtau,\grad} & R_L^{\dtau,\src} \\ 
        R^{\grad,\dtau}_L & 1 & R^{\grad,\src}_L \\ 
        R^{\src,\dtau}_L & R^{\src,\grad}_L & 1 
    \end{pmatrix} 
    \begin{pmatrix} \dtau_{LM} \\ \grad_{LM} \\ \src_{LM} \end{pmatrix}
    \notag \\ 
    &\equiv \bR{R}_L \begin{pmatrix} \dtau_{LM} \\ \grad_{LM} \\ \src_{LM} \end{pmatrix}
    \,. 
}
Motivated by the above equation, we can construct a set of 
``bias-hardened" estimators for reconstructing $\dtau$, $\grad$ and $\src$: 
\al{
    \begin{pmatrix} \estt^{\rm BH}_{LM}  \\ \estp^{\rm BH}_{LM} 
    \\ \ests^{\rm BH}_{LM} \end{pmatrix} 
    = \bR{R}^{-1}_L
    \begin{pmatrix} \estt_{LM} \\ \estp_{LM} \\ \ests_{LM} \end{pmatrix}    
    \,. 
}
The above estimators are unbiased by construction. 
If we only consider mitigation of the lensing contribution in estimating $\tau$, 
the ``lensing-hardened" $\tau$ estimator becomes \cite{Namikawa:2012:bhe}:
\al{
    \estt^{\rm BH,lens}_{LM} = \frac{\estt_{LM} 
    - R_L^{\dtau,\grad}\estp_{LM}}{1-R_L^{\dtau,\grad}R_L^{\grad,\dtau}}
    \,. 
}
The above estimators simultaneously estimate $\tau$ and lensing (and point source), 
and looses small amount of signal-to-noise compared to 
the standard quadratic estimator. 
The reconstruction noise level of $\estt^{\rm BH,lens}_{LM}$ is increased 
by $\sim 0-40\%$ depending on scales compared to $\estt_{LM}$. 
The noise levels of $\estt^{\rm BH}_{LM}$ and $\estt^{\rm BH,lens}_{LM}$ 
are almost the same. 

\section{Chi-squared PTE of measured spectra}

\begin{table}
\bc
\caption{The PTE values for our measured cross-spectrum, $C_L^{\dtau y}$, with variation of analysis choices.}
\begin{tabular}{lc}\\
 & PTE \\ \hline\hline
Baseline & $0.18$ \\ \hline
NILC for CMB map & $0.58$ \\ 
MILCA for $y$-map & $0.22$ \\ 
G50 mask for $y$-map & $0.76$ \\ 
No SZ CMB map & $0.28$ \\ \hline
\end{tabular}
\label{table:pte}
\ec
\end{table}

Here we show the $\chi^2$-PTE for the measured spectrum. 
The $\chi^2$ for the null spectrum is defined by (e.g. \cite{B2I}): 
\al{
    \chi^2 = \sum_{bb'} \hC^{\tau y}_b\{\bR{Cov}^{-1}\}_{bb'}\hC^{\tau y}_{b'} 
    \,, 
}
where $b,b'$ are the index of the multipole bin, $\hC^{\tau y}_b$ is 
the measured cross spectrum, and $\bR{Cov}$ is the band-power covariance of the cross spectrum. 
The covariance is evaluated from the simulation. 
We evaluate the PTE of the observed $\chi^2$ value against 
the distribution of the simulation. 

Table \ref{table:pte} summarizes the $\chi^2$-PTE values for the measured spectra shown in Fig.~\ref{fig:compsep}. From these values, all of the cross-spectra are consistent with null within $2\sigma$. 

\section{Effect of temperature - density relation on the cross spectrum between optical-depth fluctuations and y-map}
\label{app:Te}

In the main analysis, we ignored the fluctuations of the gas temperature inside the ionized bubbles. In this section, we discuss the impact of the gas temperature fluctuations on $\Cty$. We use the standard parametrization of the temperature-density relation, $T_{\rm b}=T_{\rm b,0}(1+\delta_{\rm g})^\beta$, where $T_{\rm b,0}$ is the temperature at the mean gas density, $\delta_{\rm g}$ is the gas density fluctuations and $\beta$ is the power-law index with $\beta<1$ \cite{Hui1997}. Then, \eq{eq:y_reio} is modified as:
\begin{equation}
    y(\uvec{n}) = \frac{\sigma_{\rm T} n_{p,0}}{m_{\rm e}c^2}
    \int d\chi a k_{\rm B} T_{\rm b,0}[1+\delta_{\rm g}(\uvec{n},\chi)]^\beta x_e (\uvec{n},\chi) 
    \,, \label{eq:y_reo_app}
\end{equation}
To simplify the above equation, we make the following approximations: 1) $T_{\rm b,0}$ does not depend on spatial positions and temperature fluctuations are induced solely by the gas density fluctuations, 2) the gas density fluctuations are small so that the higher-order terms, $\mC{O}(\delta_{\rm g}^2)$, are negligible, and 3) $\ave{\delta_{\rm g}x_{\rm e}x_{\rm e}}$ is equivalent to $\ave{x_{\rm e}}P_{\delta_{\rm g}x_{\rm e}}$, which is explicitly expressed in Eq.(15) of \cite{Mortonson:2006:model}. 
Then, we can rewrite the modified version of $\Cty$ as:
\al{
    \Cty &= \frac{k_{\rm B}\sigma_{\rm T}^2n_{\rm p0}^2}{m_{\rm e}c^2}\int \frac{d\chi} {a^4\chi^2}T_{\rm b,0}(\chi)P_{x_{\rm e}x_{\rm e}}\left(k=\frac{L+1/2}{\chi},\chi\right) 
    \notag \\
    &+ \frac{k_{\rm B}\sigma_{\rm T}^2n_{\rm p0}^2}{m_{\rm e}c^2}\int \frac{d\chi} {a^4\chi^2}T_{\rm b,0}(\chi)\beta x_{\rm e} P_{\delta_{\rm g}x_{\rm e}}\left(k=\frac{L+1/2}{\chi},\chi\right)
    \,. \label{eq:cltauy_mod}
}
The first term is the same as the $\dtau$-$y$ cross-angular power spectrum defined in \eq{eq:cltauy} and the second term is an additional term due to the inclusion of the temperature-density relation. 

\begin{figure}[t]
\centering
\includegraphics[width=85mm,clip]{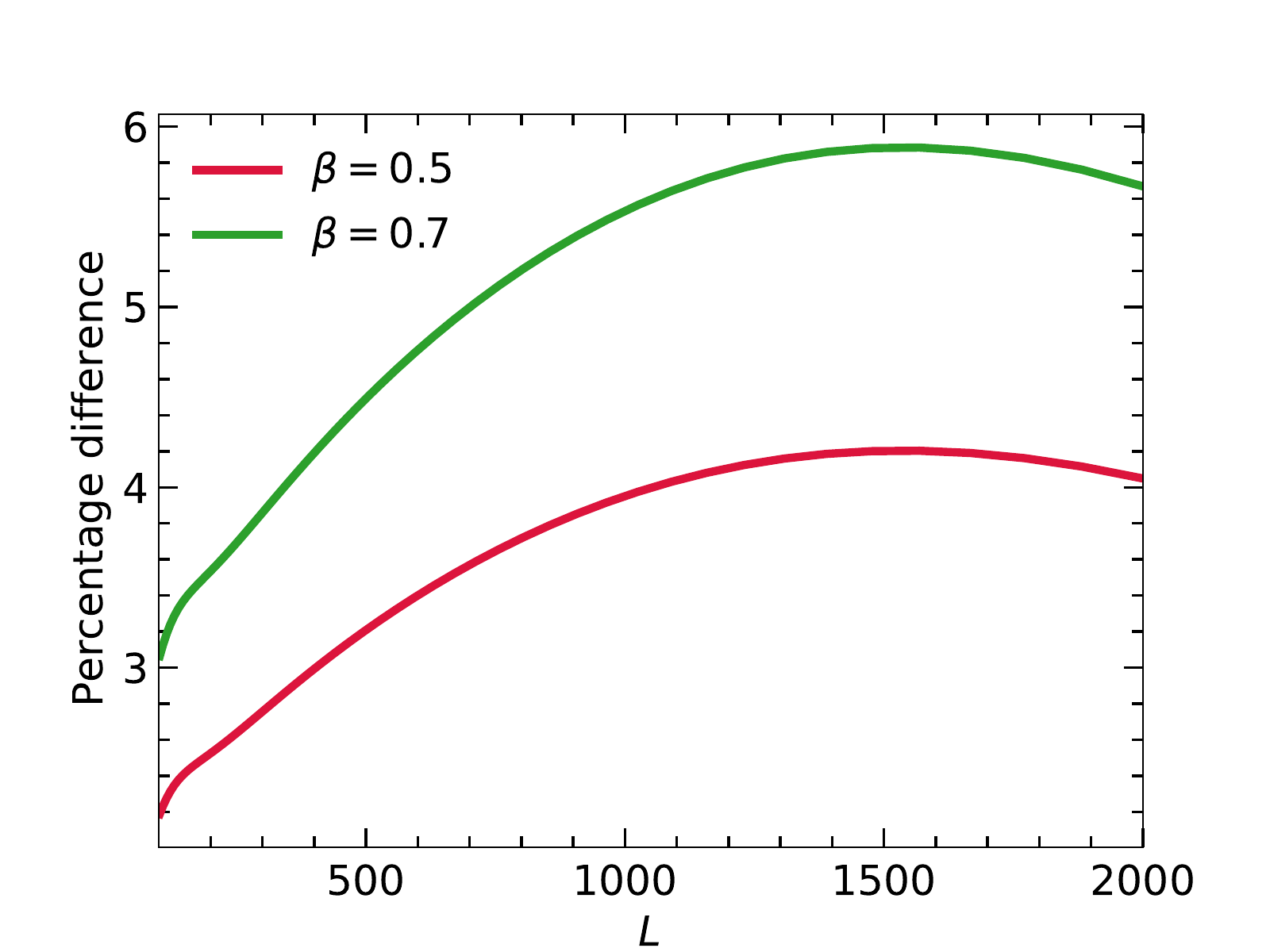}
\caption{
We show the percentage increase of $\Cty$ by the additional contribution in \eq{eq:cltauy_mod} which arises from the inclusion of temperature density relation, $T_{\rm b}=T_{\rm b,0}(1+\delta)^\beta$. We assume a characteristic ionized bubble radius $R_{\rm b}=5$\,Mpc and distribution width $\sigma_{\rm lnr}=\ln(2)$, and the temperature of ionized bubbles $T_{\rm b}=10^5$\,K. We show two cases, $\beta=0.5$ (red) and $\beta=0.7$ (green). 
}
\label{fig:density_correction}
\end{figure}

Fig.~\ref{fig:density_correction} shows the fractional difference of $\Cty$ with and without the additional correction in \eq{eq:cltauy_mod}. The percentage difference of the amplitude of $\Cty$ is $\sim4\%$ and $\sim6\%$ for $\beta=0.5$ and $0.7$, respectively. 
Note that we made several approximations and \eq{eq:cltauy_mod} is not valid if the perturbations of gas density become highly nonlinear. At such scales, $T_{\rm b}\simeq(\delta_{\rm g})^\beta$ and $\Cty$ contains nonlinear correlations of $\ave{(\delta_{\rm g})^\beta x_{\rm e}x_{\rm e}}$. If the fluctuations of $x_{\rm e}$ are enhanced at over-dense regions of gas, the nonlinear correlation could further enhance $\Cty$ and the constraint on $T_{\rm b}$ could be further tighten, although we need radiative transfer simulation to accurately model $\Cty$ in the presence of the gas temperature fluctuations.

\bibliographystyle{apsrev4-2}
\bibliography{cite}

\begin{thebibliography}{71}%
\makeatletter
\providecommand \@ifxundefined [1]{%
 \@ifx{#1\undefined}
}%
\providecommand \@ifnum [1]{%
 \ifnum #1\expandafter \@firstoftwo
 \else \expandafter \@secondoftwo
 \fi
}%
\providecommand \@ifx [1]{%
 \ifx #1\expandafter \@firstoftwo
 \else \expandafter \@secondoftwo
 \fi
}%
\providecommand \natexlab [1]{#1}%
\providecommand \enquote  [1]{``#1''}%
\providecommand \bibnamefont  [1]{#1}%
\providecommand \bibfnamefont [1]{#1}%
\providecommand \citenamefont [1]{#1}%
\providecommand \href@noop [0]{\@secondoftwo}%
\providecommand \href [0]{\begingroup \@sanitize@url \@href}%
\providecommand \@href[1]{\@@startlink{#1}\@@href}%
\providecommand \@@href[1]{\endgroup#1\@@endlink}%
\providecommand \@sanitize@url [0]{\catcode `\\12\catcode `\$12\catcode
  `\&12\catcode `\#12\catcode `\^12\catcode `\_12\catcode `\%12\relax}%
\providecommand \@@startlink[1]{}%
\providecommand \@@endlink[0]{}%
\providecommand \url  [0]{\begingroup\@sanitize@url \@url }%
\providecommand \@url [1]{\endgroup\@href {#1}{\urlprefix }}%
\providecommand \urlprefix  [0]{URL }%
\providecommand \Eprint [0]{\href }%
\providecommand \doibase [0]{https://doi.org/}%
\providecommand \selectlanguage [0]{\@gobble}%
\providecommand \bibinfo  [0]{\@secondoftwo}%
\providecommand \bibfield  [0]{\@secondoftwo}%
\providecommand \translation [1]{[#1]}%
\providecommand \BibitemOpen [0]{}%
\providecommand \bibitemStop [0]{}%
\providecommand \bibitemNoStop [0]{.\EOS\space}%
\providecommand \EOS [0]{\spacefactor3000\relax}%
\providecommand \BibitemShut  [1]{\csname bibitem#1\endcsname}%
\let\auto@bib@innerbib\@empty
\bibitem [{\citenamefont {McGreer}\ \emph {et~al.}(2015)\citenamefont
  {McGreer}, \citenamefont {Mesinger},\ and\ \citenamefont
  {D'Odorico}}]{McGreer:2014}%
  \BibitemOpen
  \bibfield  {author} {\bibinfo {author} {\bibfnamefont {I.}~\bibnamefont
  {McGreer}}, \bibinfo {author} {\bibfnamefont {A.}~\bibnamefont {Mesinger}},\
  and\ \bibinfo {author} {\bibfnamefont {V.}~\bibnamefont {D'Odorico}},\ }\href
  {https://doi.org/10.1093/mnras/stu2449} {\bibfield  {journal} {\bibinfo
  {journal} {\mnras}\ }\textbf {\bibinfo {volume} {447}},\ \bibinfo {pages}
  {499} (\bibinfo {year} {2015})},\ \Eprint {https://arxiv.org/abs/1411.5375}
  {1411.5375} \BibitemShut {NoStop}%
\bibitem [{\citenamefont {{Gunn}}\ and\ \citenamefont
  {{Peterson}}(1965)}]{GunnPeterson:1965}%
  \BibitemOpen
  \bibfield  {author} {\bibinfo {author} {\bibfnamefont {J.~E.}\ \bibnamefont
  {{Gunn}}}\ and\ \bibinfo {author} {\bibfnamefont {B.~A.}\ \bibnamefont
  {{Peterson}}},\ }\href {https://doi.org/10.1086/148444} {\bibfield  {journal}
  {\bibinfo  {journal} {\apj}\ }\textbf {\bibinfo {volume} {142}},\ \bibinfo
  {pages} {1633} (\bibinfo {year} {1965})}\BibitemShut {NoStop}%
\bibitem [{\citenamefont {Kulkarni}\ \emph {et~al.}(2019)\citenamefont
  {Kulkarni}, \citenamefont {Keating}, \citenamefont {Haehnelt}, \citenamefont
  {Bosman}, \citenamefont {Puchwein}, \citenamefont {Chardin},\ and\
  \citenamefont {Aubert}}]{Kulkarni:2018}%
  \BibitemOpen
  \bibfield  {author} {\bibinfo {author} {\bibfnamefont {G.}~\bibnamefont
  {Kulkarni}}, \bibinfo {author} {\bibfnamefont {L.~C.}\ \bibnamefont
  {Keating}}, \bibinfo {author} {\bibfnamefont {M.~G.}\ \bibnamefont
  {Haehnelt}}, \bibinfo {author} {\bibfnamefont {S.~E.}\ \bibnamefont
  {Bosman}}, \bibinfo {author} {\bibfnamefont {E.}~\bibnamefont {Puchwein}},
  \bibinfo {author} {\bibfnamefont {J.}~\bibnamefont {Chardin}},\ and\ \bibinfo
  {author} {\bibfnamefont {D.}~\bibnamefont {Aubert}},\ }\href
  {https://doi.org/10.1093/mnrasl/slz025} {\bibfield  {journal} {\bibinfo
  {journal} {\mnras}\ }\textbf {\bibinfo {volume} {485}},\ \bibinfo {pages}
  {L24} (\bibinfo {year} {2019})},\ \Eprint {https://arxiv.org/abs/1809.06374}
  {1809.06374} \BibitemShut {NoStop}%
\bibitem [{\citenamefont {{\textit{Planck}
  Collaboration}}(2016{\natexlab{a}})}]{P16:reion}%
  \BibitemOpen
  \bibfield  {author} {\bibinfo {author} {\bibnamefont {{\textit{Planck}
  Collaboration}}},\ }\href@noop {} {\bibfield  {journal} {\bibinfo  {journal}
  {\aap}\ }\textbf {\bibinfo {volume} {596}},\ \bibinfo {pages} {A108}
  (\bibinfo {year} {2016}{\natexlab{a}})},\ \Eprint
  {https://arxiv.org/abs/1605.03507} {1605.03507} \BibitemShut {NoStop}%
\bibitem [{\citenamefont {{\textit{Planck} Collaboration}}(2020)}]{P18:main}%
  \BibitemOpen
  \bibfield  {author} {\bibinfo {author} {\bibnamefont {{\textit{Planck}
  Collaboration}}},\ }\href {https://doi.org/10.1051/0004-6361/201833910}
  {\bibfield  {journal} {\bibinfo  {journal} {\aap}\ }\textbf {\bibinfo
  {volume} {641}},\ \bibinfo {pages} {A6} (\bibinfo {year} {2020})},\ \Eprint
  {https://arxiv.org/abs/1807.06209} {1807.06209} \BibitemShut {NoStop}%
\bibitem [{\citenamefont {McQuinn}(2016)}]{McQuinn:2016}%
  \BibitemOpen
  \bibfield  {author} {\bibinfo {author} {\bibfnamefont {M.}~\bibnamefont
  {McQuinn}},\ }\href {https://doi.org/10.1146/annurev-astro-082214-122355}
  {\bibfield  {journal} {\bibinfo  {journal} {Ann. Rev. Astron. Astrophys.}\
  }\textbf {\bibinfo {volume} {54}},\ \bibinfo {pages} {313} (\bibinfo {year}
  {2016})},\ \Eprint {https://arxiv.org/abs/1512.00086} {1512.00086}
  \BibitemShut {NoStop}%
\bibitem [{\citenamefont {{Wang}}\ and\ \citenamefont
  {{Hu}}(2006)}]{Wang:2006}%
  \BibitemOpen
  \bibfield  {author} {\bibinfo {author} {\bibfnamefont {X.}~\bibnamefont
  {{Wang}}}\ and\ \bibinfo {author} {\bibfnamefont {W.}~\bibnamefont {{Hu}}},\
  }\href {https://doi.org/10.1086/503095} {\bibfield  {journal} {\bibinfo
  {journal} {\apj}\ }\textbf {\bibinfo {volume} {643}},\ \bibinfo {pages} {585}
  (\bibinfo {year} {2006})},\ \Eprint {https://arxiv.org/abs/astro-ph/0511141}
  {astro-ph/0511141} \BibitemShut {NoStop}%
\bibitem [{\citenamefont {Dvorkin}\ and\ \citenamefont
  {Smith}(2009)}]{Dvorkin:2008:tau-est}%
  \BibitemOpen
  \bibfield  {author} {\bibinfo {author} {\bibfnamefont {C.}~\bibnamefont
  {Dvorkin}}\ and\ \bibinfo {author} {\bibfnamefont {K.~M.}\ \bibnamefont
  {Smith}},\ }\href@noop {} {\bibfield  {journal} {\bibinfo  {journal} {\prd}\
  }\textbf {\bibinfo {volume} {79}},\ \bibinfo {pages} {043003} (\bibinfo
  {year} {2009})},\ \Eprint {https://arxiv.org/abs/0812.1566} {0812.1566}
  \BibitemShut {NoStop}%
\bibitem [{\citenamefont {Gluscevic}\ \emph {et~al.}(2013)\citenamefont
  {Gluscevic}, \citenamefont {Kamionkowski},\ and\ \citenamefont
  {Hanson}}]{Gluscevic:2012:tau}%
  \BibitemOpen
  \bibfield  {author} {\bibinfo {author} {\bibfnamefont {V.}~\bibnamefont
  {Gluscevic}}, \bibinfo {author} {\bibfnamefont {M.}~\bibnamefont
  {Kamionkowski}},\ and\ \bibinfo {author} {\bibfnamefont {D.}~\bibnamefont
  {Hanson}},\ }\href@noop {} {\bibfield  {journal} {\bibinfo  {journal} {\prd}\
  }\textbf {\bibinfo {volume} {87}},\ \bibinfo {pages} {047303} (\bibinfo
  {year} {2013})},\ \Eprint {https://arxiv.org/abs/1210.5507} {1210.5507}
  \BibitemShut {NoStop}%
\bibitem [{\citenamefont {Namikawa}(2018)}]{Namikawa:2017:plktau}%
  \BibitemOpen
  \bibfield  {author} {\bibinfo {author} {\bibfnamefont {T.}~\bibnamefont
  {Namikawa}},\ }\href@noop {} {\bibfield  {journal} {\bibinfo  {journal}
  {\prd}\ }\textbf {\bibinfo {volume} {97}},\ \bibinfo {pages} {063505}
  (\bibinfo {year} {2018})},\ \Eprint {https://arxiv.org/abs/1711.00058}
  {1711.00058} \BibitemShut {NoStop}%
\bibitem [{\citenamefont {Feng}\ and\ \citenamefont
  {Holder}(2019)}]{Feng:2018:tau-phi}%
  \BibitemOpen
  \bibfield  {author} {\bibinfo {author} {\bibfnamefont {C.}~\bibnamefont
  {Feng}}\ and\ \bibinfo {author} {\bibfnamefont {G.}~\bibnamefont {Holder}},\
  }\href {https://doi.org/10.1103/PhysRevD.99.123502} {\bibfield  {journal}
  {\bibinfo  {journal} {\prd}\ }\textbf {\bibinfo {volume} {99}},\ \bibinfo
  {pages} {123502} (\bibinfo {year} {2019})},\ \Eprint
  {https://arxiv.org/abs/1808.01592} {1808.01592} \BibitemShut {NoStop}%
\bibitem [{\citenamefont {Furlanetto}\ \emph {et~al.}(2006)\citenamefont
  {Furlanetto}, \citenamefont {McQuinn},\ and\ \citenamefont
  {Hernquist}}]{Furlanetto:2006}%
  \BibitemOpen
  \bibfield  {author} {\bibinfo {author} {\bibfnamefont {S.}~\bibnamefont
  {Furlanetto}}, \bibinfo {author} {\bibfnamefont {M.}~\bibnamefont
  {McQuinn}},\ and\ \bibinfo {author} {\bibfnamefont {L.}~\bibnamefont
  {Hernquist}},\ }\href@noop {} {\bibfield  {journal} {\bibinfo  {journal}
  {\mnras}\ }\textbf {\bibinfo {volume} {365}},\ \bibinfo {pages} {115}
  (\bibinfo {year} {2006})},\ \Eprint {https://arxiv.org/abs/astro-ph/0507524}
  {astro-ph/0507524} \BibitemShut {NoStop}%
\bibitem [{\citenamefont {Zahn}\ \emph {et~al.}(2007)\citenamefont {Zahn},
  \citenamefont {Lidz}, \citenamefont {McQuinn}, \citenamefont {Dutta},
  \citenamefont {Hernquist}, \citenamefont {Zaldarriaga},\ and\ \citenamefont
  {Furlanetto}}]{Zahn:2006b}%
  \BibitemOpen
  \bibfield  {author} {\bibinfo {author} {\bibfnamefont {O.}~\bibnamefont
  {Zahn}}, \bibinfo {author} {\bibfnamefont {A.}~\bibnamefont {Lidz}}, \bibinfo
  {author} {\bibfnamefont {M.}~\bibnamefont {McQuinn}}, \bibinfo {author}
  {\bibfnamefont {S.}~\bibnamefont {Dutta}}, \bibinfo {author} {\bibfnamefont
  {L.}~\bibnamefont {Hernquist}}, \bibinfo {author} {\bibfnamefont
  {M.}~\bibnamefont {Zaldarriaga}},\ and\ \bibinfo {author} {\bibfnamefont
  {S.~R.}\ \bibnamefont {Furlanetto}},\ }\href {https://doi.org/10.1086/509597}
  {\bibfield  {journal} {\bibinfo  {journal} {\apj}\ }\textbf {\bibinfo
  {volume} {654}},\ \bibinfo {pages} {12} (\bibinfo {year} {2007})},\ \Eprint
  {https://arxiv.org/abs/astro-ph/0604177} {astro-ph/0604177} \BibitemShut
  {NoStop}%
\bibitem [{\citenamefont {{Battaglia}}\ \emph
  {et~al.}(2013{\natexlab{a}})\citenamefont {{Battaglia}}, \citenamefont
  {{Trac}}, \citenamefont {{Cen}},\ and\ \citenamefont
  {{Loeb}}}]{Battaglia:2013}%
  \BibitemOpen
  \bibfield  {author} {\bibinfo {author} {\bibfnamefont {N.}~\bibnamefont
  {{Battaglia}}}, \bibinfo {author} {\bibfnamefont {H.}~\bibnamefont {{Trac}}},
  \bibinfo {author} {\bibfnamefont {R.}~\bibnamefont {{Cen}}},\ and\ \bibinfo
  {author} {\bibfnamefont {A.}~\bibnamefont {{Loeb}}},\ }\href
  {https://doi.org/10.1088/0004-637X/776/2/81} {\bibfield  {journal} {\bibinfo
  {journal} {\apj}\ }\textbf {\bibinfo {volume} {776}},\ \bibinfo {eid} {81}
  (\bibinfo {year} {2013}{\natexlab{a}})},\ \Eprint
  {https://arxiv.org/abs/1211.2821} {1211.2821} \BibitemShut {NoStop}%
\bibitem [{\citenamefont {Wyithe}\ and\ \citenamefont
  {Loeb}(2004)}]{Wyithe:2004}%
  \BibitemOpen
  \bibfield  {author} {\bibinfo {author} {\bibfnamefont {J.~B.}\ \bibnamefont
  {Wyithe}}\ and\ \bibinfo {author} {\bibfnamefont {A.}~\bibnamefont {Loeb}},\
  }\href {https://doi.org/10.1038/nature03033} {\bibfield  {journal} {\bibinfo
  {journal} {Nature}\ }\textbf {\bibinfo {volume} {432}},\ \bibinfo {pages}
  {194} (\bibinfo {year} {2004})},\ \Eprint
  {https://arxiv.org/abs/astro-ph/0409412} {astro-ph/0409412} \BibitemShut
  {NoStop}%
\bibitem [{\citenamefont {{Mellema}}\ \emph {et~al.}(2015)\citenamefont
  {{Mellema}}, \citenamefont {{Koopmans}}, \citenamefont {{Shukla}},
  \citenamefont {{Datta}}, \citenamefont {{Mesinger}},\ and\ \citenamefont
  {{Majumdar}}}]{Mellema:2015}%
  \BibitemOpen
  \bibfield  {author} {\bibinfo {author} {\bibfnamefont {G.}~\bibnamefont
  {{Mellema}}}, \bibinfo {author} {\bibfnamefont {L.}~\bibnamefont
  {{Koopmans}}}, \bibinfo {author} {\bibfnamefont {H.}~\bibnamefont
  {{Shukla}}}, \bibinfo {author} {\bibfnamefont {K.~K.}\ \bibnamefont
  {{Datta}}}, \bibinfo {author} {\bibfnamefont {A.}~\bibnamefont
  {{Mesinger}}},\ and\ \bibinfo {author} {\bibfnamefont {S.}~\bibnamefont
  {{Majumdar}}},\ }\href@noop {} {\bibfield  {journal} {\bibinfo  {journal}
  {Proc. Sci.}\ }\textbf {\bibinfo {volume} {AASAK14}},\ \bibinfo {eid} {10}
  (\bibinfo {year} {2015})},\ \Eprint {https://arxiv.org/abs/1501.04203}
  {1501.04203} \BibitemShut {NoStop}%
\bibitem [{\citenamefont {Bowman}\ \emph {et~al.}(2018)\citenamefont {Bowman},
  \citenamefont {Rogers}, \citenamefont {Monsalve}, \citenamefont {Mozdzen},\
  and\ \citenamefont {Mahesh}}]{Bowman:2018}%
  \BibitemOpen
  \bibfield  {author} {\bibinfo {author} {\bibfnamefont {J.~D.}\ \bibnamefont
  {Bowman}}, \bibinfo {author} {\bibfnamefont {A.~E.~E.}\ \bibnamefont
  {Rogers}}, \bibinfo {author} {\bibfnamefont {R.~A.}\ \bibnamefont
  {Monsalve}}, \bibinfo {author} {\bibfnamefont {T.~J.}\ \bibnamefont
  {Mozdzen}},\ and\ \bibinfo {author} {\bibfnamefont {N.}~\bibnamefont
  {Mahesh}},\ }\href {https://doi.org/10.1038/nature25792} {\bibfield
  {journal} {\bibinfo  {journal} {Nature}\ }\textbf {\bibinfo {volume} {555}},\
  \bibinfo {pages} {67} (\bibinfo {year} {2018})},\ \Eprint
  {https://arxiv.org/abs/1810.05912} {1810.05912} \BibitemShut {NoStop}%
\bibitem [{\citenamefont {Hill}\ \emph {et~al.}(2015)\citenamefont {Hill},
  \citenamefont {Battaglia}, \citenamefont {Chluba}, \citenamefont {Ferraro},
  \citenamefont {Schaan},\ and\ \citenamefont {Spergel}}]{Hill:2015}%
  \BibitemOpen
  \bibfield  {author} {\bibinfo {author} {\bibfnamefont {J.~C.}\ \bibnamefont
  {Hill}}, \bibinfo {author} {\bibfnamefont {N.}~\bibnamefont {Battaglia}},
  \bibinfo {author} {\bibfnamefont {J.}~\bibnamefont {Chluba}}, \bibinfo
  {author} {\bibfnamefont {S.}~\bibnamefont {Ferraro}}, \bibinfo {author}
  {\bibfnamefont {E.}~\bibnamefont {Schaan}},\ and\ \bibinfo {author}
  {\bibfnamefont {D.~N.}\ \bibnamefont {Spergel}},\ }\href
  {https://doi.org/10.1103/PhysRevLett.115.261301} {\bibfield  {journal}
  {\bibinfo  {journal} {Phys. Rev. Lett.}\ }\textbf {\bibinfo {volume} {115}},\
  \bibinfo {pages} {261301} (\bibinfo {year} {2015})},\ \Eprint
  {https://arxiv.org/abs/1507.01583} {1507.01583} \BibitemShut {NoStop}%
\bibitem [{\citenamefont {Baxter}\ \emph {et~al.}(2021)\citenamefont {Baxter},
  \citenamefont {Weinberger}, \citenamefont {Haehnelt}, \citenamefont
  {Ir\v{s}i\v{c}}, \citenamefont {Kulkarni}, \citenamefont {Pandey},\ and\
  \citenamefont {Roy}}]{Baxter:2020dvd}%
  \BibitemOpen
  \bibfield  {author} {\bibinfo {author} {\bibfnamefont {E.~J.}\ \bibnamefont
  {Baxter}}, \bibinfo {author} {\bibfnamefont {L.}~\bibnamefont {Weinberger}},
  \bibinfo {author} {\bibfnamefont {M.}~\bibnamefont {Haehnelt}}, \bibinfo
  {author} {\bibfnamefont {V.}~\bibnamefont {Ir\v{s}i\v{c}}}, \bibinfo {author}
  {\bibfnamefont {G.}~\bibnamefont {Kulkarni}}, \bibinfo {author}
  {\bibfnamefont {S.}~\bibnamefont {Pandey}},\ and\ \bibinfo {author}
  {\bibfnamefont {A.}~\bibnamefont {Roy}},\ }\href
  {https://doi.org/10.1093/mnras/stab016} {\bibfield  {journal} {\bibinfo
  {journal} {\mnras}\ }\textbf {\bibinfo {volume} {501}},\ \bibinfo {pages}
  {6215} (\bibinfo {year} {2021})},\ \Eprint {https://arxiv.org/abs/2006.09742}
  {2006.09742} \BibitemShut {NoStop}%
\bibitem [{\citenamefont {{Mesinger}}\ \emph {et~al.}(2012)\citenamefont
  {{Mesinger}}, \citenamefont {{McQuinn}},\ and\ \citenamefont
  {{Spergel}}}]{Mesinger2012}%
  \BibitemOpen
  \bibfield  {author} {\bibinfo {author} {\bibfnamefont {A.}~\bibnamefont
  {{Mesinger}}}, \bibinfo {author} {\bibfnamefont {M.}~\bibnamefont
  {{McQuinn}}},\ and\ \bibinfo {author} {\bibfnamefont {D.~N.}\ \bibnamefont
  {{Spergel}}},\ }\href {https://doi.org/10.1111/j.1365-2966.2012.20713.x}
  {\bibfield  {journal} {\bibinfo  {journal} {\mnras}\ }\textbf {\bibinfo
  {volume} {422}},\ \bibinfo {pages} {1403} (\bibinfo {year} {2012})},\ \Eprint
  {https://arxiv.org/abs/1112.1820} {arXiv:1112.1820 [astro-ph.CO]}
  \BibitemShut {NoStop}%
\bibitem [{\citenamefont {{Battaglia}}\ \emph
  {et~al.}(2013{\natexlab{b}})\citenamefont {{Battaglia}}, \citenamefont
  {{Natarajan}}, \citenamefont {{Trac}}, \citenamefont {{Cen}},\ and\
  \citenamefont {{Loeb}}}]{Battaglia2013b}%
  \BibitemOpen
  \bibfield  {author} {\bibinfo {author} {\bibfnamefont {N.}~\bibnamefont
  {{Battaglia}}}, \bibinfo {author} {\bibfnamefont {A.}~\bibnamefont
  {{Natarajan}}}, \bibinfo {author} {\bibfnamefont {H.}~\bibnamefont {{Trac}}},
  \bibinfo {author} {\bibfnamefont {R.}~\bibnamefont {{Cen}}},\ and\ \bibinfo
  {author} {\bibfnamefont {A.}~\bibnamefont {{Loeb}}},\ }\href
  {https://doi.org/10.1088/0004-637X/776/2/83} {\bibfield  {journal} {\bibinfo
  {journal} {\apj}\ }\textbf {\bibinfo {volume} {776}},\ \bibinfo {eid} {83}
  (\bibinfo {year} {2013}{\natexlab{b}})},\ \Eprint
  {https://arxiv.org/abs/1211.2832} {1211.2832} \BibitemShut {NoStop}%
\bibitem [{\citenamefont {Smith}\ and\ \citenamefont
  {Ferraro}(2017)}]{Smith:2016lnt}%
  \BibitemOpen
  \bibfield  {author} {\bibinfo {author} {\bibfnamefont {K.~M.}\ \bibnamefont
  {Smith}}\ and\ \bibinfo {author} {\bibfnamefont {S.}~\bibnamefont
  {Ferraro}},\ }\href@noop {} {\bibfield  {journal} {\bibinfo  {journal}
  {\prl}\ }\textbf {\bibinfo {volume} {119}},\ \bibinfo {pages} {021301}
  (\bibinfo {year} {2017})},\ \Eprint {https://arxiv.org/abs/1607.01769}
  {1607.01769} \BibitemShut {NoStop}%
\bibitem [{\citenamefont {Alvarez}\ \emph {et~al.}(2021)\citenamefont
  {Alvarez}, \citenamefont {Ferraro}, \citenamefont {Hill}, \citenamefont
  {Hlo\v{z}ek},\ and\ \citenamefont {Ikape}}]{Alvarez:2020}%
  \BibitemOpen
  \bibfield  {author} {\bibinfo {author} {\bibfnamefont {M.~A.}\ \bibnamefont
  {Alvarez}}, \bibinfo {author} {\bibfnamefont {S.}~\bibnamefont {Ferraro}},
  \bibinfo {author} {\bibfnamefont {J.~C.}\ \bibnamefont {Hill}}, \bibinfo
  {author} {\bibfnamefont {R.}~\bibnamefont {Hlo\v{z}ek}},\ and\ \bibinfo
  {author} {\bibfnamefont {M.}~\bibnamefont {Ikape}},\ }\href
  {https://doi.org/10.1103/PhysRevD.103.063518} {\bibfield  {journal} {\bibinfo
   {journal} {\prd}\ }\textbf {\bibinfo {volume} {103}},\ \bibinfo {pages}
  {063518} (\bibinfo {year} {2021})},\ \Eprint
  {https://arxiv.org/abs/2006.06594} {2006.06594} \BibitemShut {NoStop}%
\bibitem [{\citenamefont {De~Zotti}\ \emph {et~al.}(2016)\citenamefont
  {De~Zotti}, \citenamefont {Negrello}, \citenamefont {Castex}, \citenamefont
  {Lapi},\ and\ \citenamefont {Bonato}}]{DeZotti:2015}%
  \BibitemOpen
  \bibfield  {author} {\bibinfo {author} {\bibfnamefont {G.}~\bibnamefont
  {De~Zotti}}, \bibinfo {author} {\bibfnamefont {M.}~\bibnamefont {Negrello}},
  \bibinfo {author} {\bibfnamefont {G.}~\bibnamefont {Castex}}, \bibinfo
  {author} {\bibfnamefont {A.}~\bibnamefont {Lapi}},\ and\ \bibinfo {author}
  {\bibfnamefont {M.}~\bibnamefont {Bonato}},\ }\href
  {https://doi.org/10.1088/1475-7516/2016/03/047} {\bibfield  {journal}
  {\bibinfo  {journal} {\jcap}\ }\textbf {\bibinfo {volume} {03}},\ \bibinfo
  {pages} {047} (\bibinfo {year} {2016})},\ \Eprint
  {https://arxiv.org/abs/1512.04816} {1512.04816} \BibitemShut {NoStop}%
\bibitem [{\citenamefont {Mortonson}\ and\ \citenamefont
  {Hu}(2007)}]{Mortonson:2006:model}%
  \BibitemOpen
  \bibfield  {author} {\bibinfo {author} {\bibfnamefont {M.~J.}\ \bibnamefont
  {Mortonson}}\ and\ \bibinfo {author} {\bibfnamefont {W.}~\bibnamefont {Hu}},\
  }\href {https://doi.org/10.1086/510574} {\bibfield  {journal} {\bibinfo
  {journal} {\apj}\ }\textbf {\bibinfo {volume} {657}},\ \bibinfo {pages} {1}
  (\bibinfo {year} {2007})},\ \Eprint {https://arxiv.org/abs/astro-ph/0607652}
  {astro-ph/0607652} \BibitemShut {NoStop}%
\bibitem [{\citenamefont {Roy}\ \emph {et~al.}(2018)\citenamefont {Roy},
  \citenamefont {Lapi}, \citenamefont {Spergel},\ and\ \citenamefont
  {Baccigalupi}}]{Roy:2018}%
  \BibitemOpen
  \bibfield  {author} {\bibinfo {author} {\bibfnamefont {A.}~\bibnamefont
  {Roy}}, \bibinfo {author} {\bibfnamefont {A.}~\bibnamefont {Lapi}}, \bibinfo
  {author} {\bibfnamefont {D.}~\bibnamefont {Spergel}},\ and\ \bibinfo {author}
  {\bibfnamefont {C.}~\bibnamefont {Baccigalupi}},\ }\href
  {https://doi.org/10.1088/1475-7516/2018/05/014} {\bibfield  {journal}
  {\bibinfo  {journal} {\jcap}\ }\textbf {\bibinfo {volume} {05}},\ \bibinfo
  {pages} {014} (\bibinfo {year} {2018})},\ \Eprint
  {https://arxiv.org/abs/1801.02393} {1801.02393} \BibitemShut {NoStop}%
\bibitem [{\citenamefont {Roy}\ \emph {et~al.}(2021)\citenamefont {Roy},
  \citenamefont {Kulkarni}, \citenamefont {Meerburg}, \citenamefont
  {Challinor}, \citenamefont {Baccigalupi}, \citenamefont {Lapi},\ and\
  \citenamefont {Haehnelt}}]{Roy:2020}%
  \BibitemOpen
  \bibfield  {author} {\bibinfo {author} {\bibfnamefont {A.}~\bibnamefont
  {Roy}}, \bibinfo {author} {\bibfnamefont {G.}~\bibnamefont {Kulkarni}},
  \bibinfo {author} {\bibfnamefont {P.~D.}\ \bibnamefont {Meerburg}}, \bibinfo
  {author} {\bibfnamefont {A.}~\bibnamefont {Challinor}}, \bibinfo {author}
  {\bibfnamefont {C.}~\bibnamefont {Baccigalupi}}, \bibinfo {author}
  {\bibfnamefont {A.}~\bibnamefont {Lapi}},\ and\ \bibinfo {author}
  {\bibfnamefont {M.~G.}\ \bibnamefont {Haehnelt}},\ }\href
  {https://doi.org/10.1088/1475-7516/2021/01/003} {\bibfield  {journal}
  {\bibinfo  {journal} {\jcap}\ }\textbf {\bibinfo {volume} {01}},\ \bibinfo
  {pages} {003} (\bibinfo {year} {2021})},\ \Eprint
  {https://arxiv.org/abs/2004.02927} {2004.02927} \BibitemShut {NoStop}%
\bibitem [{\citenamefont {{Remazeilles}}\ \emph {et~al.}(2011)\citenamefont
  {{Remazeilles}}, \citenamefont {{Delabrouille}},\ and\ \citenamefont
  {{Cardoso}}}]{Remazeilles:2011:NILC}%
  \BibitemOpen
  \bibfield  {author} {\bibinfo {author} {\bibfnamefont {M.}~\bibnamefont
  {{Remazeilles}}}, \bibinfo {author} {\bibfnamefont {J.}~\bibnamefont
  {{Delabrouille}}},\ and\ \bibinfo {author} {\bibfnamefont {J.-F.}\
  \bibnamefont {{Cardoso}}},\ }\href
  {https://doi.org/10.1111/j.1365-2966.2010.17624.x} {\bibfield  {journal}
  {\bibinfo  {journal} {\mnras}\ }\textbf {\bibinfo {volume} {410}},\ \bibinfo
  {pages} {2481} (\bibinfo {year} {2011})},\ \Eprint
  {https://arxiv.org/abs/1006.5599} {1006.5599} \BibitemShut {NoStop}%
\bibitem [{\citenamefont {{Hurier}}\ \emph {et~al.}(2013)\citenamefont
  {{Hurier}}, \citenamefont {{Mac{\'\i}as-P{\'e}rez}},\ and\ \citenamefont
  {{Hildebrandt}}}]{Hurier:2013:MILCA}%
  \BibitemOpen
  \bibfield  {author} {\bibinfo {author} {\bibfnamefont {G.}~\bibnamefont
  {{Hurier}}}, \bibinfo {author} {\bibfnamefont {J.~F.}\ \bibnamefont
  {{Mac{\'\i}as-P{\'e}rez}}},\ and\ \bibinfo {author} {\bibfnamefont
  {S.}~\bibnamefont {{Hildebrandt}}},\ }\href
  {https://doi.org/10.1051/0004-6361/201321891} {\bibfield  {journal} {\bibinfo
   {journal} {\aap}\ }\textbf {\bibinfo {volume} {558}},\ \bibinfo {eid} {A118}
  (\bibinfo {year} {2013})},\ \Eprint {https://arxiv.org/abs/1007.1149}
  {1007.1149} \BibitemShut {NoStop}%
\bibitem [{\citenamefont {{\textit{Planck}
  Collaboration}}(2016{\natexlab{b}})}]{P15:phi}%
  \BibitemOpen
  \bibfield  {author} {\bibinfo {author} {\bibnamefont {{\textit{Planck}
  Collaboration}}},\ }\href {https://doi.org/10.1051/0004-6361/201525941}
  {\bibfield  {journal} {\bibinfo  {journal} {\aap}\ }\textbf {\bibinfo
  {volume} {594}},\ \bibinfo {pages} {A15} (\bibinfo {year}
  {2016}{\natexlab{b}})},\ \Eprint {https://arxiv.org/abs/1502.01591}
  {1502.01591} \BibitemShut {NoStop}%
\bibitem [{\citenamefont {Namikawa}\ \emph {et~al.}(2013)\citenamefont
  {Namikawa}, \citenamefont {Hanson},\ and\ \citenamefont
  {Takahashi}}]{Namikawa:2012:bhe}%
  \BibitemOpen
  \bibfield  {author} {\bibinfo {author} {\bibfnamefont {T.}~\bibnamefont
  {Namikawa}}, \bibinfo {author} {\bibfnamefont {D.}~\bibnamefont {Hanson}},\
  and\ \bibinfo {author} {\bibfnamefont {R.}~\bibnamefont {Takahashi}},\ }\href
  {https://doi.org/10.1093/mnras/stt195} {\bibfield  {journal} {\bibinfo
  {journal} {\mnras}\ }\textbf {\bibinfo {volume} {431}},\ \bibinfo {pages}
  {609} (\bibinfo {year} {2013})},\ \Eprint {https://arxiv.org/abs/1209.0091}
  {1209.0091} \BibitemShut {NoStop}%
\bibitem [{\citenamefont {Hu}\ and\ \citenamefont
  {Okamoto}(2002)}]{HuOkamoto:2001}%
  \BibitemOpen
  \bibfield  {author} {\bibinfo {author} {\bibfnamefont {W.}~\bibnamefont
  {Hu}}\ and\ \bibinfo {author} {\bibfnamefont {T.}~\bibnamefont {Okamoto}},\
  }\href@noop {} {\bibfield  {journal} {\bibinfo  {journal} {\apj}\ }\textbf
  {\bibinfo {volume} {574}},\ \bibinfo {pages} {566} (\bibinfo {year}
  {2002})},\ \Eprint {https://arxiv.org/abs/astro-ph/0111606}
  {astro-ph/0111606} \BibitemShut {NoStop}%
\bibitem [{\citenamefont {Okamoto}\ and\ \citenamefont
  {Hu}(2003)}]{OkamotoHu:quad}%
  \BibitemOpen
  \bibfield  {author} {\bibinfo {author} {\bibfnamefont {T.}~\bibnamefont
  {Okamoto}}\ and\ \bibinfo {author} {\bibfnamefont {W.}~\bibnamefont {Hu}},\
  }\href@noop {} {\bibfield  {journal} {\bibinfo  {journal} {\prd}\ }\textbf
  {\bibinfo {volume} {67}},\ \bibinfo {pages} {083002} (\bibinfo {year}
  {2003})},\ \Eprint {https://arxiv.org/abs/astro-ph/0301031}
  {astro-ph/0301031} \BibitemShut {NoStop}%
\bibitem [{\citenamefont {Osborne}\ \emph {et~al.}(2014)\citenamefont
  {Osborne}, \citenamefont {Hanson},\ and\ \citenamefont
  {Dor\'e}}]{Osborne:2013nna}%
  \BibitemOpen
  \bibfield  {author} {\bibinfo {author} {\bibfnamefont {S.~J.}\ \bibnamefont
  {Osborne}}, \bibinfo {author} {\bibfnamefont {D.}~\bibnamefont {Hanson}},\
  and\ \bibinfo {author} {\bibfnamefont {O.}~\bibnamefont {Dor\'e}},\ }\href
  {https://doi.org/10.1088/1475-7516/2014/03/024} {\bibfield  {journal}
  {\bibinfo  {journal} {\jcap}\ }\textbf {\bibinfo {volume} {03}},\ \bibinfo
  {pages} {024} (\bibinfo {year} {2014})},\ \Eprint
  {https://arxiv.org/abs/1310.7547} {1310.7547} \BibitemShut {NoStop}%
\bibitem [{\citenamefont {Namikawa}\ and\ \citenamefont
  {Takahashi}(2014)}]{Namikawa:2013:bhepol}%
  \BibitemOpen
  \bibfield  {author} {\bibinfo {author} {\bibfnamefont {T.}~\bibnamefont
  {Namikawa}}\ and\ \bibinfo {author} {\bibfnamefont {R.}~\bibnamefont
  {Takahashi}},\ }\href@noop {} {\bibfield  {journal} {\bibinfo  {journal}
  {\mnras}\ }\textbf {\bibinfo {volume} {438}},\ \bibinfo {pages} {1507}
  (\bibinfo {year} {2014})},\ \Eprint {https://arxiv.org/abs/1310.2372}
  {1310.2372} \BibitemShut {NoStop}%
\bibitem [{\citenamefont {{\textit{Planck}
  Collaboration}}(2016{\natexlab{c}})}]{P16:tSZ}%
  \BibitemOpen
  \bibfield  {author} {\bibinfo {author} {\bibnamefont {{\textit{Planck}
  Collaboration}}},\ }\href {https://doi.org/10.1051/0004-6361/201525826}
  {\bibfield  {journal} {\bibinfo  {journal} {\aap}\ }\textbf {\bibinfo
  {volume} {594}},\ \bibinfo {eid} {A22} (\bibinfo {year}
  {2016}{\natexlab{c}})},\ \Eprint {https://arxiv.org/abs/1502.01596}
  {1502.01596} \BibitemShut {NoStop}%
\bibitem [{\citenamefont {Hill}\ and\ \citenamefont
  {Spergel}(2014)}]{Hill:2013dxa}%
  \BibitemOpen
  \bibfield  {author} {\bibinfo {author} {\bibfnamefont {J.~C.}\ \bibnamefont
  {Hill}}\ and\ \bibinfo {author} {\bibfnamefont {D.~N.}\ \bibnamefont
  {Spergel}},\ }\href@noop {} {\bibfield  {journal} {\bibinfo  {journal}
  {\jcap}\ }\textbf {\bibinfo {volume} {02}},\ \bibinfo {pages} {030} (\bibinfo
  {year} {2014})},\ \Eprint {https://arxiv.org/abs/1312.4525} {1312.4525}
  \BibitemShut {NoStop}%
\bibitem [{\citenamefont {{Battaglia}}\ \emph {et~al.}(2010)\citenamefont
  {{Battaglia}}, \citenamefont {{Bond}}, \citenamefont {{Pfrommer}},
  \citenamefont {{Sievers}},\ and\ \citenamefont {{Sijacki}}}]{Battaglia2010}%
  \BibitemOpen
  \bibfield  {author} {\bibinfo {author} {\bibfnamefont {N.}~\bibnamefont
  {{Battaglia}}}, \bibinfo {author} {\bibfnamefont {J.~R.}\ \bibnamefont
  {{Bond}}}, \bibinfo {author} {\bibfnamefont {C.}~\bibnamefont {{Pfrommer}}},
  \bibinfo {author} {\bibfnamefont {J.~L.}\ \bibnamefont {{Sievers}}},\ and\
  \bibinfo {author} {\bibfnamefont {D.}~\bibnamefont {{Sijacki}}},\ }\href
  {https://doi.org/10.1088/0004-637X/725/1/91} {\bibfield  {journal} {\bibinfo
  {journal} {\apj}\ }\textbf {\bibinfo {volume} {725}},\ \bibinfo {pages} {91}
  (\bibinfo {year} {2010})},\ \Eprint {https://arxiv.org/abs/1003.4256}
  {1003.4256} \BibitemShut {NoStop}%
\bibitem [{\citenamefont {{Foreman-Mackey}}\ \emph {et~al.}(2013)\citenamefont
  {{Foreman-Mackey}}, \citenamefont {{Hogg}}, \citenamefont {{Lang}},\ and\
  \citenamefont {{Goodman}}}]{emcee}%
  \BibitemOpen
  \bibfield  {author} {\bibinfo {author} {\bibfnamefont {D.}~\bibnamefont
  {{Foreman-Mackey}}}, \bibinfo {author} {\bibfnamefont {D.~W.}\ \bibnamefont
  {{Hogg}}}, \bibinfo {author} {\bibfnamefont {D.}~\bibnamefont {{Lang}}},\
  and\ \bibinfo {author} {\bibfnamefont {J.}~\bibnamefont {{Goodman}}},\ }\href
  {https://doi.org/10.1086/670067} {\bibfield  {journal} {\bibinfo  {journal}
  {Publications of the Astronomical Society of the Pacific}\ }\textbf {\bibinfo
  {volume} {125}},\ \bibinfo {pages} {306} (\bibinfo {year} {2013})},\ \Eprint
  {https://arxiv.org/abs/1202.3665} {1202.3665} \BibitemShut {NoStop}%
\bibitem [{\citenamefont {Davies}\ \emph {et~al.}(2018)\citenamefont {Davies}
  \emph {et~al.}}]{Davies:2018}%
  \BibitemOpen
  \bibfield  {author} {\bibinfo {author} {\bibfnamefont {F.~B.}\ \bibnamefont
  {Davies}} \emph {et~al.},\ }\href {https://doi.org/10.3847/1538-4357/aad6dc}
  {\bibfield  {journal} {\bibinfo  {journal} {\apj}\ }\textbf {\bibinfo
  {volume} {864}},\ \bibinfo {pages} {142} (\bibinfo {year} {2018})},\ \Eprint
  {https://arxiv.org/abs/1802.06066} {1802.06066} \BibitemShut {NoStop}%
\bibitem [{\citenamefont {Greig}\ \emph {et~al.}(2017)\citenamefont {Greig},
  \citenamefont {Mesinger}, \citenamefont {Haiman},\ and\ \citenamefont
  {Simcoe}}]{Greig:2016}%
  \BibitemOpen
  \bibfield  {author} {\bibinfo {author} {\bibfnamefont {B.}~\bibnamefont
  {Greig}}, \bibinfo {author} {\bibfnamefont {A.}~\bibnamefont {Mesinger}},
  \bibinfo {author} {\bibfnamefont {Z.}~\bibnamefont {Haiman}},\ and\ \bibinfo
  {author} {\bibfnamefont {R.~A.}\ \bibnamefont {Simcoe}},\ }\href
  {https://doi.org/10.1093/mnras/stw3351} {\bibfield  {journal} {\bibinfo
  {journal} {\mnras}\ }\textbf {\bibinfo {volume} {466}},\ \bibinfo {pages}
  {4239} (\bibinfo {year} {2017})},\ \Eprint {https://arxiv.org/abs/1606.00441}
  {1606.00441} \BibitemShut {NoStop}%
\bibitem [{\citenamefont {Battaglia}(2016)}]{Battaglia:2016}%
  \BibitemOpen
  \bibfield  {author} {\bibinfo {author} {\bibfnamefont {N.}~\bibnamefont
  {Battaglia}},\ }\href {https://doi.org/10.1088/1475-7516/2016/08/058}
  {\bibfield  {journal} {\bibinfo  {journal} {\jcap}\ }\textbf {\bibinfo
  {volume} {08}},\ \bibinfo {pages} {058} (\bibinfo {year} {2016})},\ \Eprint
  {https://arxiv.org/abs/1607.02442} {1607.02442} \BibitemShut {NoStop}%
\bibitem [{\citenamefont {{Battaglia}}\ \emph {et~al.}(2012)\citenamefont
  {{Battaglia}}, \citenamefont {{Bond}}, \citenamefont {{Pfrommer}},\ and\
  \citenamefont {{Sievers}}}]{Battaglia2012b}%
  \BibitemOpen
  \bibfield  {author} {\bibinfo {author} {\bibfnamefont {N.}~\bibnamefont
  {{Battaglia}}}, \bibinfo {author} {\bibfnamefont {J.~R.}\ \bibnamefont
  {{Bond}}}, \bibinfo {author} {\bibfnamefont {C.}~\bibnamefont {{Pfrommer}}},\
  and\ \bibinfo {author} {\bibfnamefont {J.~L.}\ \bibnamefont {{Sievers}}},\
  }\href {https://doi.org/10.1088/0004-637X/758/2/75} {\bibfield  {journal}
  {\bibinfo  {journal} {\apj}\ }\textbf {\bibinfo {volume} {758}},\ \bibinfo
  {eid} {75} (\bibinfo {year} {2012})},\ \Eprint
  {https://arxiv.org/abs/1109.3711} {1109.3711} \BibitemShut {NoStop}%
\bibitem [{\citenamefont {{Bolton}}\ \emph {et~al.}(2010)\citenamefont
  {{Bolton}}, \citenamefont {{Becker}}, \citenamefont {{Wyithe}}, \citenamefont
  {{Haehnelt}},\ and\ \citenamefont {{Sargent}}}]{2010MNRAS.406..612B}%
  \BibitemOpen
  \bibfield  {author} {\bibinfo {author} {\bibfnamefont {J.~S.}\ \bibnamefont
  {{Bolton}}}, \bibinfo {author} {\bibfnamefont {G.~D.}\ \bibnamefont
  {{Becker}}}, \bibinfo {author} {\bibfnamefont {J.~S.~B.}\ \bibnamefont
  {{Wyithe}}}, \bibinfo {author} {\bibfnamefont {M.~G.}\ \bibnamefont
  {{Haehnelt}}},\ and\ \bibinfo {author} {\bibfnamefont {W.~L.~W.}\
  \bibnamefont {{Sargent}}},\ }\href
  {https://doi.org/10.1111/j.1365-2966.2010.16701.x} {\bibfield  {journal}
  {\bibinfo  {journal} {\mnras}\ }\textbf {\bibinfo {volume} {406}},\ \bibinfo
  {pages} {612} (\bibinfo {year} {2010})},\ \Eprint
  {https://arxiv.org/abs/1001.3415} {arXiv:1001.3415 [astro-ph.CO]}
  \BibitemShut {NoStop}%
\bibitem [{\citenamefont {McQuinn}\ \emph {et~al.}(2009)\citenamefont
  {McQuinn}, \citenamefont {Lidz}, \citenamefont {Zaldarriaga}, \citenamefont
  {Hernquist}, \citenamefont {Hopkins}, \citenamefont {Dutta},\ and\
  \citenamefont {Faucher-Giguere}}]{McQuinn:2008am}%
  \BibitemOpen
  \bibfield  {author} {\bibinfo {author} {\bibfnamefont {M.}~\bibnamefont
  {McQuinn}}, \bibinfo {author} {\bibfnamefont {A.}~\bibnamefont {Lidz}},
  \bibinfo {author} {\bibfnamefont {M.}~\bibnamefont {Zaldarriaga}}, \bibinfo
  {author} {\bibfnamefont {L.}~\bibnamefont {Hernquist}}, \bibinfo {author}
  {\bibfnamefont {P.~F.}\ \bibnamefont {Hopkins}}, \bibinfo {author}
  {\bibfnamefont {S.}~\bibnamefont {Dutta}},\ and\ \bibinfo {author}
  {\bibfnamefont {C.~A.}\ \bibnamefont {Faucher-Giguere}},\ }\href
  {https://doi.org/10.1088/0004-637X/694/2/842} {\bibfield  {journal} {\bibinfo
   {journal} {Astrophys. J.}\ }\textbf {\bibinfo {volume} {694}},\ \bibinfo
  {pages} {842} (\bibinfo {year} {2009})},\ \Eprint
  {https://arxiv.org/abs/0807.2799} {0807.2799} \BibitemShut {NoStop}%
\bibitem [{\citenamefont {Hanany}\ \emph {et~al.}(2019)\citenamefont {Hanany}
  \emph {et~al.}}]{Hanany:2019lle}%
  \BibitemOpen
  \bibfield  {author} {\bibinfo {author} {\bibfnamefont {S.}~\bibnamefont
  {Hanany}} \emph {et~al.},\ }\href@noop {} {\  (\bibinfo {year} {2019})},\
  \Eprint {https://arxiv.org/abs/1902.10541} {1902.10541} \BibitemShut
  {NoStop}%
\bibitem [{\citenamefont {Sehgal}\ \emph {et~al.}(2019)\citenamefont {Sehgal}
  \emph {et~al.}}]{Sehgal:2019ewc}%
  \BibitemOpen
  \bibfield  {author} {\bibinfo {author} {\bibfnamefont {N.}~\bibnamefont
  {Sehgal}} \emph {et~al.},\ }\href@noop {} {\  (\bibinfo {year} {2019})},\
  \Eprint {https://arxiv.org/abs/1906.10134} {1906.10134} \BibitemShut
  {NoStop}%
\bibitem [{\citenamefont {Smith}\ \emph {et~al.}(2012)\citenamefont {Smith},
  \citenamefont {Hanson}, \citenamefont {LoVerde}, \citenamefont {Hirata},\
  and\ \citenamefont {Zahn}}]{Smith:2010gu}%
  \BibitemOpen
  \bibfield  {author} {\bibinfo {author} {\bibfnamefont {K.~M.}\ \bibnamefont
  {Smith}}, \bibinfo {author} {\bibfnamefont {D.}~\bibnamefont {Hanson}},
  \bibinfo {author} {\bibfnamefont {M.}~\bibnamefont {LoVerde}}, \bibinfo
  {author} {\bibfnamefont {C.~M.}\ \bibnamefont {Hirata}},\ and\ \bibinfo
  {author} {\bibfnamefont {O.}~\bibnamefont {Zahn}},\ }\href@noop {} {\bibfield
   {journal} {\bibinfo  {journal} {\jcap}\ }\textbf {\bibinfo {volume} {06}},\
  \bibinfo {pages} {014} (\bibinfo {year} {2012})},\ \Eprint
  {https://arxiv.org/abs/1010.0048} {1010.0048} \BibitemShut {NoStop}%
\bibitem [{\citenamefont {Guzman}\ and\ \citenamefont
  {Meyers}(2021)}]{Guzman:2021}%
  \BibitemOpen
  \bibfield  {author} {\bibinfo {author} {\bibfnamefont {E.}~\bibnamefont
  {Guzman}}\ and\ \bibinfo {author} {\bibfnamefont {J.}~\bibnamefont
  {Meyers}},\ }\href@noop {} {\  (\bibinfo {year} {2021})},\ \Eprint
  {https://arxiv.org/abs/2101.01214} {2101.01214} \BibitemShut {NoStop}%
\bibitem [{\citenamefont {Schaan}\ \emph {et~al.}(2021)\citenamefont {Schaan}
  \emph {et~al.}}]{Schaan2020}%
  \BibitemOpen
  \bibfield  {author} {\bibinfo {author} {\bibfnamefont {E.}~\bibnamefont
  {Schaan}} \emph {et~al.},\ }\href
  {https://doi.org/10.1103/PhysRevD.103.063513} {\bibfield  {journal} {\bibinfo
   {journal} {\prd}\ }\textbf {\bibinfo {volume} {103}},\ \bibinfo {pages}
  {063513} (\bibinfo {year} {2021})},\ \Eprint
  {https://arxiv.org/abs/2009.05557} {2009.05557} \BibitemShut {NoStop}%
\bibitem [{\citenamefont {Amodeo}\ \emph {et~al.}(2021)\citenamefont {Amodeo}
  \emph {et~al.}}]{Amodeo2020}%
  \BibitemOpen
  \bibfield  {author} {\bibinfo {author} {\bibfnamefont {S.}~\bibnamefont
  {Amodeo}} \emph {et~al.},\ }\href
  {https://doi.org/10.1103/PhysRevD.103.063514} {\bibfield  {journal} {\bibinfo
   {journal} {\prd}\ }\textbf {\bibinfo {volume} {103}},\ \bibinfo {pages}
  {063514} (\bibinfo {year} {2021})},\ \Eprint
  {https://arxiv.org/abs/2009.05558} {2009.05558} \BibitemShut {NoStop}%
\bibitem [{\citenamefont {Sailer}\ \emph {et~al.}(2020)\citenamefont {Sailer},
  \citenamefont {Schaan},\ and\ \citenamefont {Ferraro}}]{Sailer:2020:bhe}%
  \BibitemOpen
  \bibfield  {author} {\bibinfo {author} {\bibfnamefont {N.}~\bibnamefont
  {Sailer}}, \bibinfo {author} {\bibfnamefont {E.}~\bibnamefont {Schaan}},\
  and\ \bibinfo {author} {\bibfnamefont {S.}~\bibnamefont {Ferraro}},\ }\href
  {https://doi.org/10.1103/PhysRevD.102.063517} {\bibfield  {journal} {\bibinfo
   {journal} {\prd}\ }\textbf {\bibinfo {volume} {102}},\ \bibinfo {pages}
  {063517} (\bibinfo {year} {2020})},\ \Eprint
  {https://arxiv.org/abs/2007.04325} {2007.04325} \BibitemShut {NoStop}%
\bibitem [{\citenamefont {Lidz}\ and\ \citenamefont
  {Malloy}(2014)}]{Lidz:2014jxa}%
  \BibitemOpen
  \bibfield  {author} {\bibinfo {author} {\bibfnamefont {A.}~\bibnamefont
  {Lidz}}\ and\ \bibinfo {author} {\bibfnamefont {M.}~\bibnamefont {Malloy}},\
  }\href {https://doi.org/10.1088/0004-637X/788/2/175} {\bibfield  {journal}
  {\bibinfo  {journal} {Astrophys. J.}\ }\textbf {\bibinfo {volume} {788}},\
  \bibinfo {pages} {175} (\bibinfo {year} {2014})},\ \Eprint
  {https://arxiv.org/abs/1403.6350} {1403.6350} \BibitemShut {NoStop}%
\bibitem [{\citenamefont {D'Aloisio}\ \emph {et~al.}(2019)\citenamefont
  {D'Aloisio}, \citenamefont {McQuinn}, \citenamefont {Maupin}, \citenamefont
  {Davies}, \citenamefont {Trac}, \citenamefont {Fuller},\ and\ \citenamefont
  {Upton~Sanderbeck}}]{DAloisio:2018rzi}%
  \BibitemOpen
  \bibfield  {author} {\bibinfo {author} {\bibfnamefont {A.}~\bibnamefont
  {D'Aloisio}}, \bibinfo {author} {\bibfnamefont {M.}~\bibnamefont {McQuinn}},
  \bibinfo {author} {\bibfnamefont {O.}~\bibnamefont {Maupin}}, \bibinfo
  {author} {\bibfnamefont {F.~B.}\ \bibnamefont {Davies}}, \bibinfo {author}
  {\bibfnamefont {H.}~\bibnamefont {Trac}}, \bibinfo {author} {\bibfnamefont
  {S.}~\bibnamefont {Fuller}},\ and\ \bibinfo {author} {\bibfnamefont {P.~R.}\
  \bibnamefont {Upton~Sanderbeck}},\ }\href
  {https://doi.org/10.3847/1538-4357/ab0d83} {\bibfield  {journal} {\bibinfo
  {journal} {Astrophys. J.}\ }\textbf {\bibinfo {volume} {874}},\ \bibinfo
  {pages} {154} (\bibinfo {year} {2019})},\ \Eprint
  {https://arxiv.org/abs/1807.09282} {1807.09282} \BibitemShut {NoStop}%
\bibitem [{\citenamefont {{McQuinn}}(2012)}]{Mcquinn2012}%
  \BibitemOpen
  \bibfield  {author} {\bibinfo {author} {\bibfnamefont {M.}~\bibnamefont
  {{McQuinn}}},\ }\href {https://doi.org/10.1111/j.1365-2966.2012.21792.x}
  {\bibfield  {journal} {\bibinfo  {journal} {\mnras}\ }\textbf {\bibinfo
  {volume} {426}},\ \bibinfo {pages} {1349} (\bibinfo {year} {2012})},\ \Eprint
  {https://arxiv.org/abs/1206.1335} {1206.1335} \BibitemShut {NoStop}%
\bibitem [{\citenamefont {Finlator}\ \emph {et~al.}(2018)\citenamefont
  {Finlator}, \citenamefont {Keating}, \citenamefont {Oppenheimer},
  \citenamefont {Dav\'e},\ and\ \citenamefont {Zackrisson}}]{Finlator:2018lbl}%
  \BibitemOpen
  \bibfield  {author} {\bibinfo {author} {\bibfnamefont {K.}~\bibnamefont
  {Finlator}}, \bibinfo {author} {\bibfnamefont {L.}~\bibnamefont {Keating}},
  \bibinfo {author} {\bibfnamefont {B.~D.}\ \bibnamefont {Oppenheimer}},
  \bibinfo {author} {\bibfnamefont {R.}~\bibnamefont {Dav\'e}},\ and\ \bibinfo
  {author} {\bibfnamefont {E.}~\bibnamefont {Zackrisson}},\ }\href
  {https://doi.org/10.1093/mnras/sty1949} {\bibfield  {journal} {\bibinfo
  {journal} {Mon. Not. Roy. Astron. Soc.}\ }\textbf {\bibinfo {volume} {480}},\
  \bibinfo {pages} {2628} (\bibinfo {year} {2018})},\ \Eprint
  {https://arxiv.org/abs/1805.00099} {1805.00099} \BibitemShut {NoStop}%
\bibitem [{\citenamefont {Gaikwad}\ \emph {et~al.}(2020)\citenamefont {Gaikwad}
  \emph {et~al.}}]{Gaikwad:2020art}%
  \BibitemOpen
  \bibfield  {author} {\bibinfo {author} {\bibfnamefont {P.}~\bibnamefont
  {Gaikwad}} \emph {et~al.},\ }\href {https://doi.org/10.1093/mnras/staa907}
  {\bibfield  {journal} {\bibinfo  {journal} {\mnras}\ }\textbf {\bibinfo
  {volume} {494}},\ \bibinfo {pages} {5091} (\bibinfo {year} {2020})},\ \Eprint
  {https://arxiv.org/abs/2001.10018} {2001.10018} \BibitemShut {NoStop}%
\bibitem [{\citenamefont {Bolton}\ \emph {et~al.}(2010)\citenamefont {Bolton},
  \citenamefont {Becker}, \citenamefont {Wyithe}, \citenamefont {Haehnelt},\
  and\ \citenamefont {Sargent}}]{Bolton:2010gr}%
  \BibitemOpen
  \bibfield  {author} {\bibinfo {author} {\bibfnamefont {J.~S.}\ \bibnamefont
  {Bolton}}, \bibinfo {author} {\bibfnamefont {G.~D.}\ \bibnamefont {Becker}},
  \bibinfo {author} {\bibfnamefont {J.~B.}\ \bibnamefont {Wyithe}}, \bibinfo
  {author} {\bibfnamefont {M.~G.}\ \bibnamefont {Haehnelt}},\ and\ \bibinfo
  {author} {\bibfnamefont {W.~L.}\ \bibnamefont {Sargent}},\ }\href
  {https://doi.org/10.1111/j.1365-2966.2010.16701.x} {\bibfield  {journal}
  {\bibinfo  {journal} {\mnras}\ }\textbf {\bibinfo {volume} {406}},\ \bibinfo
  {pages} {612} (\bibinfo {year} {2010})},\ \Eprint
  {https://arxiv.org/abs/1001.3415} {1001.3415} \BibitemShut {NoStop}%
\bibitem [{\citenamefont {Katz}\ \emph {et~al.}(2019)\citenamefont {Katz},
  \citenamefont {Kimm}, \citenamefont {Haehnelt}, \citenamefont {Sijacki},
  \citenamefont {Rosdahl},\ and\ \citenamefont {Blaizot}}]{Katz:2018xle}%
  \BibitemOpen
  \bibfield  {author} {\bibinfo {author} {\bibfnamefont {H.}~\bibnamefont
  {Katz}}, \bibinfo {author} {\bibfnamefont {T.}~\bibnamefont {Kimm}}, \bibinfo
  {author} {\bibfnamefont {M.~G.}\ \bibnamefont {Haehnelt}}, \bibinfo {author}
  {\bibfnamefont {D.}~\bibnamefont {Sijacki}}, \bibinfo {author} {\bibfnamefont
  {J.}~\bibnamefont {Rosdahl}},\ and\ \bibinfo {author} {\bibfnamefont
  {J.}~\bibnamefont {Blaizot}},\ }\href {https://doi.org/10.1093/mnras/sty3154}
  {\bibfield  {journal} {\bibinfo  {journal} {\mnras}\ }\textbf {\bibinfo
  {volume} {483}},\ \bibinfo {pages} {1029} (\bibinfo {year} {2019})},\ \Eprint
  {https://arxiv.org/abs/1806.07474} {1806.07474} \BibitemShut {NoStop}%
\bibitem [{\citenamefont {Tumlinson}\ and\ \citenamefont
  {Shull}(2000)}]{Tumlinson:1999iu}%
  \BibitemOpen
  \bibfield  {author} {\bibinfo {author} {\bibfnamefont {J.}~\bibnamefont
  {Tumlinson}}\ and\ \bibinfo {author} {\bibfnamefont {J.}~\bibnamefont
  {Shull}},\ }\href {https://doi.org/10.1086/312432} {\bibfield  {journal}
  {\bibinfo  {journal} {Astrophys. J. Lett.}\ }\textbf {\bibinfo {volume}
  {528}},\ \bibinfo {pages} {L65} (\bibinfo {year} {2000})},\ \Eprint
  {https://arxiv.org/abs/astro-ph/9911339} {astro-ph/9911339} \BibitemShut
  {NoStop}%
\bibitem [{\citenamefont {Pagano}\ \emph {et~al.}(2020)\citenamefont {Pagano},
  \citenamefont {Delouis}, \citenamefont {Mottet}, \citenamefont {Puget},\ and\
  \citenamefont {Vibert}}]{Pagano:2019}%
  \BibitemOpen
  \bibfield  {author} {\bibinfo {author} {\bibfnamefont {L.}~\bibnamefont
  {Pagano}}, \bibinfo {author} {\bibfnamefont {J.-M.}\ \bibnamefont {Delouis}},
  \bibinfo {author} {\bibfnamefont {S.}~\bibnamefont {Mottet}}, \bibinfo
  {author} {\bibfnamefont {J.-L.}\ \bibnamefont {Puget}},\ and\ \bibinfo
  {author} {\bibfnamefont {L.}~\bibnamefont {Vibert}},\ }\href
  {https://doi.org/10.1051/0004-6361/201936630} {\bibfield  {journal} {\bibinfo
   {journal} {\aap}\ }\textbf {\bibinfo {volume} {635}},\ \bibinfo {pages}
  {A99} (\bibinfo {year} {2020})},\ \Eprint {https://arxiv.org/abs/1908.09856}
  {1908.09856} \BibitemShut {NoStop}%
\bibitem [{\citenamefont {Reichardt}\ \emph {et~al.}(2020)\citenamefont
  {Reichardt} \emph {et~al.}}]{Reichardt:2020}%
  \BibitemOpen
  \bibfield  {author} {\bibinfo {author} {\bibfnamefont {C.}~\bibnamefont
  {Reichardt}} \emph {et~al.} (\bibinfo {collaboration} {SPT}),\ }\href@noop {}
  {\bibfield  {journal} {\bibinfo  {journal} {arXiv e-prints}\ } (\bibinfo
  {year} {2020})},\ \Eprint {https://arxiv.org/abs/2002.06197} {2002.06197}
  \BibitemShut {NoStop}%
\bibitem [{\citenamefont {Henderson}\ \emph {et~al.}(2016)\citenamefont
  {Henderson} \emph {et~al.}}]{Henderson:2015}%
  \BibitemOpen
  \bibfield  {author} {\bibinfo {author} {\bibfnamefont {S.}~\bibnamefont
  {Henderson}} \emph {et~al.},\ }\href
  {https://doi.org/10.1007/s10909-016-1575-z} {\bibfield  {journal} {\bibinfo
  {journal} {J. Low Temp. Phys.}\ }\textbf {\bibinfo {volume} {184}},\ \bibinfo
  {pages} {772} (\bibinfo {year} {2016})},\ \Eprint
  {https://arxiv.org/abs/1510.02809} {1510.02809} \BibitemShut {NoStop}%
\bibitem [{\citenamefont {Benson}\ \emph {et~al.}(2014)\citenamefont {Benson}
  \emph {et~al.}}]{Benson:2014}%
  \BibitemOpen
  \bibfield  {author} {\bibinfo {author} {\bibfnamefont {B.}~\bibnamefont
  {Benson}} \emph {et~al.} (\bibinfo {collaboration} {SPT-3G}),\ }\href
  {https://doi.org/10.1117/12.2057305} {\bibfield  {journal} {\bibinfo
  {journal} {Proc. SPIE Int. Soc. Opt. Eng.}\ }\textbf {\bibinfo {volume}
  {9153}},\ \bibinfo {pages} {91531P} (\bibinfo {year} {2014})},\ \Eprint
  {https://arxiv.org/abs/1407.2973} {1407.2973} \BibitemShut {NoStop}%
\bibitem [{Sim(2019)}]{SimonsObservatory}%
  \BibitemOpen
  \href@noop {} {\emph {\bibinfo {title} {{The Simons Observatory: Astro2020
  Decadal Project Whitepaper}}}},\ Vol.~\bibinfo {volume} {51}\ (\bibinfo
  {year} {2019})\ \Eprint {https://arxiv.org/abs/1907.08284} {1907.08284}
  \BibitemShut {NoStop}%
\bibitem [{CMB(2019)}]{CMBS4}%
  \BibitemOpen
  \href@noop {} {\emph {\bibinfo {title} {CMB-S4 Science Case, Reference
  Design, and Project Plan}}}\ (\bibinfo {year} {2019})\ \Eprint
  {https://arxiv.org/abs/1907.04473} {1907.04473} \BibitemShut {NoStop}%
\bibitem [{\citenamefont {Feng}\ and\ \citenamefont
  {Holder}(2018)}]{Feng:2018:3pt}%
  \BibitemOpen
  \bibfield  {author} {\bibinfo {author} {\bibfnamefont {C.}~\bibnamefont
  {Feng}}\ and\ \bibinfo {author} {\bibfnamefont {G.}~\bibnamefont {Holder}},\
  }\href {https://doi.org/10.1103/PhysRevD.97.123523} {\bibfield  {journal}
  {\bibinfo  {journal} {\prd}\ }\textbf {\bibinfo {volume} {97}},\ \bibinfo
  {pages} {123523} (\bibinfo {year} {2018})},\ \Eprint
  {https://arxiv.org/abs/1801.05396} {1801.05396} \BibitemShut {NoStop}%
\bibitem [{\citenamefont {{Hazumi}}\ \emph {et~al.}(2019)\citenamefont
  {{Hazumi}} \emph {et~al.}}]{LiteBIRD}%
  \BibitemOpen
  \bibfield  {author} {\bibinfo {author} {\bibfnamefont {M.}~\bibnamefont
  {{Hazumi}}} \emph {et~al.},\ }\href
  {https://doi.org/10.1007/s10909-019-02150-5} {\bibfield  {journal} {\bibinfo
  {journal} {J. Low. Temp. Phys.}\ }\textbf {\bibinfo {volume} {194}},\
  \bibinfo {pages} {443} (\bibinfo {year} {2019})}\BibitemShut {NoStop}%
\bibitem [{\citenamefont {Smith}\ \emph {et~al.}(2018)\citenamefont {Smith},
  \citenamefont {Madhavacheril}, \citenamefont {M\"unchmeyer}, \citenamefont
  {Ferraro}, \citenamefont {Giri},\ and\ \citenamefont
  {Johnson}}]{Smith:2018:kSZ}%
  \BibitemOpen
  \bibfield  {author} {\bibinfo {author} {\bibfnamefont {K.~M.}\ \bibnamefont
  {Smith}}, \bibinfo {author} {\bibfnamefont {M.~S.}\ \bibnamefont
  {Madhavacheril}}, \bibinfo {author} {\bibfnamefont {M.}~\bibnamefont
  {M\"unchmeyer}}, \bibinfo {author} {\bibfnamefont {S.}~\bibnamefont
  {Ferraro}}, \bibinfo {author} {\bibfnamefont {U.}~\bibnamefont {Giri}},\ and\
  \bibinfo {author} {\bibfnamefont {M.~C.}\ \bibnamefont {Johnson}},\
  }\href@noop {} {\  (\bibinfo {year} {2018})},\ \Eprint
  {https://arxiv.org/abs/1810.13423} {1810.13423} \BibitemShut {NoStop}%
\bibitem [{\citenamefont {{\textsc{Bicep2} Collaboration}}(2014)}]{B2I}%
  \BibitemOpen
  \bibfield  {author} {\bibinfo {author} {\bibnamefont {{\textsc{Bicep2}
  Collaboration}}},\ }\href {https://doi.org/10.1103/PhysRevLett.112.241101}
  {\bibfield  {journal} {\bibinfo  {journal} {\prl}\ }\textbf {\bibinfo
  {volume} {112}},\ \bibinfo {pages} {241101} (\bibinfo {year} {2014})},\
  \Eprint {https://arxiv.org/abs/1403.3985} {1403.3985} \BibitemShut {NoStop}%
\bibitem [{\citenamefont {{Hui}}\ and\ \citenamefont
  {{Gnedin}}(1997)}]{Hui1997}%
  \BibitemOpen
  \bibfield  {author} {\bibinfo {author} {\bibfnamefont {L.}~\bibnamefont
  {{Hui}}}\ and\ \bibinfo {author} {\bibfnamefont {N.~Y.}\ \bibnamefont
  {{Gnedin}}},\ }\href {https://doi.org/10.1093/mnras/292.1.27} {\bibfield
  {journal} {\bibinfo  {journal} {\mnras}\ }\textbf {\bibinfo {volume} {292}},\
  \bibinfo {pages} {27} (\bibinfo {year} {1997})},\ \Eprint
  {https://arxiv.org/abs/astro-ph/9612232} {astro-ph/9612232} \BibitemShut
  {NoStop}%
\end{thebibliography}%

\end{document}